\documentclass[aps,prd,groupedaddress,showpacs,showkeys]{revtex4} 
\usepackage{amssymb} 
\usepackage{amsmath} 
\usepackage{amsfonts}
\usepackage{epsfig}

\newcommand{\seq}{\begin{subequations}}
\newcommand{\sen}{\end{subequations}}
\newcommand{\eq}{\begin{eqnarray}}
\newcommand{\en}{\end{eqnarray}}
\newcommand{\bea}{\begin{eqnarray}}
\newcommand{\eea}{\end{eqnarray}}

\newcommand{\ra}{\rangle}
\newcommand{\la}{\langle}
\newcommand{\mnz}{\stackrel{\!\!\!\!\!\circ}{m_N}}
\newcommand{\gaz}{\stackrel{\!\!\!\!\!\circ}{g_A}}

\begin{document}

\title{Chiral corrections to the vector and axial couplings of quarks and
baryons}

\noindent
\author{Amand Faessler$^{1}$,   
        Thomas Gutsche$^{1}$,
        Barry R. Holstein$^{2}$,
        Valery E. Lyubovitskij$^1$\footnote{On leave of absence
        from Department of Physics, Tomsk State University,
        634050 Tomsk, Russia}
\vspace*{1.2\baselineskip}}

\affiliation{$^1$ Institut f\"ur Theoretische Physik,
Universit\"at T\"ubingen,\\
Auf der Morgenstelle 14, D--72076 T\"ubingen, Germany
\vspace*{1.2\baselineskip} \\
$^2$ Department of Physics--LGRT, University of
Massachusetts, Amherst, MA 01003 USA\\}

\date{\today}

\begin{abstract}

We calculate chiral corrections to the semileptonic vector and axial
quark coupling constants using a manifestly Lorentz covariant chiral
quark approach up to order ${\cal O}(p^4)$ in the two-- and tree--flavor 
picture.  These couplings are then used in the evaluation of the 
corresponding couplings which govern the semileptonic transitions 
between octet baryon states. In the calculation of baryon matrix 
elements we use a general ansatz for the spatial form of the quark 
wave function, without referring to a specific realization of 
hadronization and confinement of quarks in baryons. 
Matching the physical amplitudes calculated within our approach to 
the model--independent predictions of baryon chiral perturbation 
theory (ChPT) allows to deduce a connection between 
our parameters and those of baryon ChPT.

\end{abstract}



\maketitle

\newpage

\section{Introduction}

The study of the semileptonic decays of the baryon octet $B_i \to
B_j e \bar\nu_e$ presents an opportunity to shed light on the
Cabibbo--Kobayashi--Maskawa (CKM) matrix element $V_{us}$.  At zero
momentum transfer, the weak baryon matrix elements for the $B_i \to B_j
e \bar\nu_e$ transitions are determined by just two constants --- 
the vector coupling $g_V^{B_iB_j}$ and its axial counterpart
$g_A^{B_iB_j}$. In the limit of exact SU(3) symmetry $g_V^{B_iB_j}$
and $g_A^{B_iB_j}$ are expressed in terms of basic parameters --- the
vector couplings are given in terms of well--known Clebsch--Gordan
coefficients which are fixed due to the conservation of the vector
current (CVC), while the axial couplings are given in terms of the
simple SU(3) octet axial--vector couplings $F$ and $D$.  The
Ademollo--Gatto theorem (AGT)~\cite{Ademollo:1964sr} protects the
vector form factors from leading SU(3)--breaking corrections
generated by the mass difference of strange and nonstrange quarks.
The first nonvanishing breaking effects start at second order in
symmetry--breaking.  As stressed in Ref.~\cite{Cabibbo:2003ea} this
vanishing of the first--order correction to the vector hyperon form
factors $g_V^{B_iB_j}$ presents an opportunity to determine $V_{us}$
from the direct measurement of $V_{us} g_V^{B_iB_j}$.  The axial
form factors, on the other hand, contain symmetry--breaking
corrections already at first order.  Note that the experimental data
on baryon semileptonic decays~\cite{Yao:2006px} are well described
by Cabibbo theory~\cite{Cabibbo:1963yz}, which assumes SU(3)
invariance of strong interactions.  However, for a precise
determination of $V_{us}$ one needs to include leading and perhaps
subleading SU(3) breaking corrections.

The theoretical analysis of SU(3) breaking corrections to hyperon
semileptonic decay form factors has been performed in various
approaches~\cite{Donoghue:1992dd}--\cite{Lacour:2007wm}, including
quark and soliton models, $1/N_c$ expansion of QCD, chiral
perturbation theory (ChPT), lattice QCD, {\it etc.}  Quark models,
in particular, have had a major impact on the understanding of the
phenomenology of hyperon semileptonic decays.  The original
predictions of the naive SU(6)--model~\cite{Close:1979bt} have been
substantially improved by inclusion of
relativistic~\cite{Chodos:1974pn} and SU(3) symmetry breaking
effects~\cite{Donoghue:1981uk,Kubodera:1984qd,Schlumpf:1994fb},
gluon~\cite{Hogaasen:1987nj} and meson--cloud
corrections~\cite{Kubodera:1984qd,Thomas:1982kv}. However, a fully
consistent presentation of chiral corrections (both SU(3)--symmetric
and SU(3)--breaking) to semileptonic form factors of baryons is still
awaited, although a model--independent inclusion of some chiral
corrections to semileptonic form factors of baryons has been
performed using different versions of the chiral effective theory of
baryons (including baryon ChPT and heavy baryon
ChPT)~\cite{Krause:1990xc}--\cite{Lacour:2007wm}. 
Recently a complete calculation of the SU(3)--breaking corrections to the 
hyperon vector form factors up to ${\cal O}(p^4)$ in covariant baryon 
ChPT has been presented in~\cite{Lacour:2007wm}. Note, that a
detailed ChPT analysis of the nucleon axial coupling/form factor has
also been performed in Refs.~\cite{Kambor:1998pi}--\cite{Schindler:2006it}.  

In the present paper we evaluate chiral corrections to the
semileptonic vector and axial quark coupling constants, using a
manifestly Lorentz covariant chiral quark approach up to 
order ${\cal O}(p^4)$ in the two-- and tree--flavor picture.
Here SU(3) breaking corrections are generated by the mass difference
of strange ($s$) and nonstrange ($u$, $d$) current quarks. We
proceed as follows.  First, we calculate the vector and axial {\it
quark} couplings including chiral effects.  Then we use the weak
quark transition operators containing these couplings to evaluate
the octet baryon matrix elements involved in the semileptonic
transitions.  Performing the matching of the baryon matrix elements
to those derived in baryon ChPT we deduce relations between the 
parameters of the two approaches.  This matching guarantees inclusion 
of chiral corrections to baryon observables, which is consistent with
QCD. In the calculation of the baryon matrix elements we employ a
general ansatz for the spatial form of the quark wave function, without
referring to any specific realization of hadronization and
confinement of quarks in baryons. In a forthcoming  
paper~\cite{Faessler:2007qm} we will consider the evaluation of the 
baryon matrix elements within a specific Lorentz and gauge invariant 
quark model~\cite{Ivanov:1996pz} explicitly including the internal 
quark dynamics. Note that in Refs.~\cite{Faessler:2005gd,Faessler:2006ky} 
we performed an analogous study of the electromagnetic properties of
the baryon octet and the $\Delta(1230)$--resonance. In particular, 
we developed an approach based on a nonlinear chirally symmetric
Lagrangian, which involves constituent quarks {\it and} chiral
fields.  In a first step, this Lagrangian was used to dress the
constituent quarks by a cloud of light pseudoscalar mesons and other
(virtual) heavy states using the calculational technique of infrared
dimensional regularization (IDR)~\cite{Becher:1999he}. Then within a
formal chiral expansion, we calculated the dressed transition
operators relevant for the interaction of the quarks with external
fields in the presence of a virtual meson cloud. In a following
step, these dressed operators were used to calculate baryon matrix
elements including internal dynamics of valence quarks  
(Note that a simpler and more phenomenological quark
model based on similar ideas regarding the dressing of
constituent quarks by the meson cloud has been developed in
Refs.~\cite{PCQM}.) 
We treat the constituent quarks as the intermediate degrees of freedom 
between the current quarks (building blocks of the QCD Lagrangian) 
and the hadrons (building blocks of ChPT). This concept dates back 
to the pioneering works of Refs.~\cite{Altarelli:1973ff,Manohar:1983md}. 
Furthermore, our strategy in dressing the constituent quarks by a 
cloud of pseudoscalar mesons is motivated by the procedure pursued 
in Ref.~\cite{Manohar:1983md}. Recent analyses of experiments at 
Jefferson Lab (TJLAB), Fermilab, BNL and IHEP (Protvino) 
renewed the interest in the concept of constituent quarks. The obtained 
data can be interpreted in a picture, where the hadronic quasiparticle 
substructure is assumed to consist of constituent quarks with nontrivial 
form factors. These experiments also initiated new progress in the 
manifestation of constituent degrees of freedom in hadron phenomenology 
(see, e.g. Refs.~\cite{Petronzio:2003bw}). 

The present approach has the intrinsic advantage that it is a priori 
not restricted to small energy or momentum transfers. In a full evaluation,
when taking into account the effects of the internal dynamics of
valence quarks, one can describe hadron form factors at much higher
Euclidean momentum squared when compared to ChPT. 
When restricting to the inclusion of valence 
quark effects our approach was successfully 
applied to different problems of light baryons and also heavy baryons 
containing a one, two and three heavy quarks 
(see Refs.~\cite{Ivanov:1996pz,Faessler:2006ky}). We achieved good 
agreement with existing data and gave certain predictions for
future experiments. E.g. our predictions for the semileptonic,  
nonleptonic and strong decays of heavy-light baryons were later 
confirmed experimentally. On the other hand, in 
Refs.~\cite{Faessler:2005gd,Faessler:2006ky}
we developed the formalism in order to include chiral effects in a way 
consistent with low-energy theorems and the infrared structure of QCD. 
Consistency in the present formalism with ChPT is limited since we 
cannot consider baryonic matrix elements in Minkowski space. 
Also, our approach does not provide any constraints for 
the expansion parameter of standard ChPT. 

In the present manuscript we proceed as follows. First, in
Section~II, we discuss the basic notions of our approach, which is
in direct line to our previous work of 
Refs.~\cite{Faessler:2005gd,Faessler:2006ky}. That is, we derive 
a chiral Lagrangian motivated by baryon
ChPT~\cite{Becher:1999he,Gasser:1987rb}, and formulate it in terms
of quark and mesonic degrees of freedom. Using constituent quarks
dressed with a cloud of light pseudoscalar mesons and other heavy
states, we derive dressed transition operators within the chiral
expansion, which are in turn used in quark model to produce baryon 
matrix elements. In Section~III we derive specific expressions for 
the vector and axial baryon semileptonic decay constants, while in 
Section~IV we give the numerical analysis of the axial nucleon charge 
and the vector and axial hyperon semileptonic couplings.  Finally, 
in Section~V we present a short summary of our results. 

\section{Approach}

\subsection{Chiral Lagrangian}
\label{Chir_Lagrangian}

\noindent

The SU(3) chiral quark Lagrangian ${\cal L}_{qU}$ 
(up to ${\cal O}(p^3)$), which dynamically generates dressing of 
the constituent quarks by mesonic degrees of freedom, consists of 
three primary pieces ${\cal L}_{q}$, ${\cal L}_{qq}$ and 
${\cal L}_{U}$: 
\eq\label{L_qU}
{\cal L}_{qU} \,
= \, {\cal L}_{q} + {\cal L}_{qq} + {\cal L}_{U}\,, \hspace*{.5cm} 
{\cal L}_q \, =
\, {\cal L}^{(1)}_q + {\cal L}^{(2)}_q + {\cal L}^{(3)}_q +
\cdots\,, \hspace*{.5cm} 
{\cal L}_{qq} \, = \, {\cal L}^{(3)}_{qq} 
+ \cdots\,, \hspace*{.5cm} 
{\cal L}_{U} \, = \, {\cal L}_{U}^{(2)} +
\cdots\,.
\en
The superscript $(i)$ attached to ${\cal L}^{(i)}_{U}$  
and ${\cal L}^{(i)}_{q(qq)}$ 
denotes the low energy dimension of the Lagrangian:
\seq\label{L_exp}
\eq
{\cal L}_{U}^{(2)} &=&\frac{F^2}{4} \la{u_\mu u^\mu
+ \chi_+}\ra\,, \hspace*{.5cm} {\cal L}^{(1)}_q \, = \,  \bar q
\left[ i \, \slash\!\!\!\! D - m + \frac{1}{2} \, g \, \slash\!\!\!
u \, \gamma^5 \right] q\,,
\label{L_exp1} \\[2mm]
{\cal L}^{(2)}_q & = & \frac{C_3^q}{2} \, \la{u_\mu u^\mu}\ra \,
\bar q \, q \, + \, \frac{C_4^q}{4}\, \bar q\, i\, \sigma^{\mu\nu}\,
[u_\mu, u_\nu]\, q \, + \, \cdots \,, \label{L_exp2}\\[2mm]
{\cal L}^{(3)}_q &=& \frac{D_{16}^q}{2} \,\bar q \, \slash\!\!\! u
\, \gamma^5 \, q \la \, \chi_+ \ra + \frac{D_{17}^q}{8} \,\bar q \, \{
\slash\!\!\! u \, \gamma^5 \,, \hat\chi_+ \} \, q \, + \, \cdots \,,
\label{L_exp3}\\[2mm] 
{\cal L}^{(3)}_{qq} &=& \frac{D_{1}^{qq}}{2} \,\bar q \, \slash\!\!\! u
\, \gamma^5 q \ \bar q q \, \la \chi_+ \ra 
+ \frac{D_{2}^{qq}}{8} \,\bar q \, \{ \slash\!\!\! u \, \gamma^5 \,, 
\hat\chi_+ \} q \  \bar q q 
+ \frac{D_{3}^{qq}}{2} \,\bar q \, \slash\!\!\! u \, \gamma^5 q 
\ \bar q \, \hat\chi_+ \, q \, + \, \cdots \,,\label{L_exp4} 
\en 
\sen 
where the symbols $\la \,\, \ra$, $[ \,\, ]$ and  $\{ \,\, \}$ 
occurring in Eq.~(\ref{L_exp}) denote the trace over   
flavor matrices, commutator, and anticommutator, respectively.    
In Eq.~(\ref{L_exp}) we display only the terms involved in the  
calculation of vector and axial quark/baryon coupling constants.   
In addition to the one--body quark Lagrangian we included also the   
two--body part ${\cal L}_{qq}$. The detailed form of the chiral Lagrangian  
used in the calculations of electromagnetic properties of baryons  
can be found in Refs.~\cite{Faessler:2005gd,Faessler:2006ky}.  

The Lagrangians (\ref{L_exp}) contain the basic building blocks of
our approach.  The couplings $m$ and $g$ denote the quark mass and
axial charge in the chiral limit ({\it i.e.}, they are counted as
quantities of order ${\cal O}(1)$ in the chiral expansion), $q$ is
the triplet of $u,d,s$--quark fields, $C_i^q$ and $D_i^q$ are the
second-- and third--order one--body quark low--energy coupling 
constants (LEC's), respectively, while $D_i^{qq}$ are the 
third--order two--body quark LEC's. 
The LEC's encode the (virtual) contributions due to heavy states. 
We denote the SU(3) quark LEC's by capital letters in order
to distinguish them from the SU(2) LEC's $c_i^q$ and $d_i^q$. Also,  
for the one--body quark LEC's we use the additional 
superscript ``$q$'' to
differentiate them from the analogs ChPT LEC's: $C_i$, $D_i$ in
SU(3) and $c_i$, $d_i$ in SU(2). For the two--body quark LEC's we 
attach the superscript ``$qq$''. The octet of pseudoscalar fields
\eq \phi = \sum_{i=1}^{8} \phi_i\lambda_i = \sqrt{2} \left(
\begin{array}{ccc}
\pi^0/\sqrt{2} + \eta/\sqrt{6}\,\, & \,\, \pi^+ \,\, & \, K^+ \\
\pi^- \,\, & \,\, -\pi^0/\sqrt{2}+\eta/\sqrt{6}\,\, & \, K^0\\
K^-\,\, & \,\, \bar K^0 \,\, & \, -2\eta/\sqrt{6}\\
\end{array}
\right)
\en
is contained in the SU(3) matrix
$U = u^2 = {\rm exp}(i\phi/F)$ where $F$ is the octet decay constant.
We use the standard notations~\cite{Gasser:1987rb,Becher:1999he}
\eq\label{ChPT_basics}
& &D_\mu = \partial_\mu + \Gamma_\mu, \hspace*{.3cm}
\Gamma_\mu = \frac{1}{2} [u^\dagger, \partial_\mu u]
- \frac{i}{2} u^\dagger r_\mu u
- \frac{i}{2} u l_\mu u^\dagger\,, \hspace*{.3cm}
u_\mu = i \{ u^\dagger,  \partial_\mu u \} +
u^\dagger r_\mu u - u l_\mu u^\dagger, \nonumber\\
& &\chi_\pm = u^\dagger \chi u^\dagger \pm u \chi^\dagger u,
\hspace*{.3cm} \chi = 2 B {\cal M} + \cdots \,, \hspace*{.3cm}
\hat\chi_+ = \chi_+ - \frac{1}{3} \la{\chi_+}\ra\,.
\en
The fields $r_\mu$ and $l_\mu$ include external vector $(v_\mu)$ and
axial $(a_\mu)$ fields: $r_\mu \, = \, v_\mu + a_\mu \,, \hspace*{.2cm}
l_\mu \, = \, v_\mu - a_\mu \,.$ ${\cal M} = {\rm diag}\{\hat m,
\hat m, \hat m_s\}$ is the mass matrix of current quarks (we work in
the isospin symmetry limit with $\hat m_{u}= \hat m_{d}=\hat{m}=7$~MeV 
and the mass of the strange quark $\hat m_s$ is related to the
nonstrange one via $\hat m_s \simeq 25 \, \hat m$). The quark vacuum
condensate parameter is denoted by $B = - \la 0|\bar u u|0 \ra/F^2 =
     - \la 0|\bar d d|0 \ra/F^2\,.$
To distinguish between constituent and current quark masses we
attach the symbol $\ {\bf\hat{}}$ ("hat") when referring to the
current quark masses. We rely on the standard picture of chiral
symmetry breaking~($B \gg F$). To leading order in the chiral
expansion the masses of pseudoscalar mesons are given by
$M_{\pi}^2=2 \hat m B, \hspace*{.25cm} M_{K}^2=(\hat m + \hat m_s)
B, \hspace*{.25cm} M_{\eta}^2= \frac{2}{3} (\hat m + 2 \hat m_s)
B\,.$  For the numerical analysis we will use: 
$M_{\pi} = 139.57$ MeV, $M_K = 493.677$ MeV (the charged pion and 
kaon masses), $M_\eta = 547.51$ MeV, $F = F_\pi = 92.4$ MeV 
in SU(2) and $F = (F_\pi + F_K)/2$ in SU(3) with 
$F_K/F_\pi = 1.22$~\cite{Gasser:1984gg}.

Reduction of the SU(3) Lagrangian to its SU(2) counterpart is
straightforward. The quark triplet $(u,d,s)$ and meson octet are
replaced by the quark doublet $(u,d)$ and the pion triplet, 
respectively. Likewise, the LEC's $C_i^q$, $D_i^q$ and $D_i^{qq}$ 
should be replaced by their SU(2) analogue $c_i^q$, $d_i^q$ 
and $d_i^{qq}$. Note that the SU(2) Lagrangian does {\it not} 
contain the LEC's $d_{17}^q$ and $d_{2(3)}^{qq}$. 
Also, we should use ${\cal M} = {\rm diag}\{\hat m, \hat m\}$ and 
$\hat\chi_+ = \chi_+ - \frac{1}{2} \la{\chi_+}\ra$ instead of 
the corresponding quantities defined in Eq.~(\ref{ChPT_basics}).

\subsection{Dressing of quark operators}

Any bare quark operator can be dressed by a cloud of pseudoscalar 
mesons and heavy states in a straightforward manner by use of the 
effective chirally--invariant Lagrangian ${\cal L}_{qU}$. 
In Refs.~\cite{Faessler:2005gd,Faessler:2006ky} we illustrated the 
technique of dressing in the case of the electromagnetic quark 
operator and performed a detailed analysis of the electromagnetic 
properties of the baryon octet and of the
$\Delta \to N \gamma$ transition. Here we extend our method to the
case of vector and axial quark operators. First, we define the bare
vector and axial quark transition operators constructed from quark 
fields of flavor $i$ and $j$ as: 
\eq\label{bare_V}
& &J_{\mu, \, V}(q) = \int d^4x \, e^{-iqx} \, j_{\mu, \, V}(x) \,, \quad
\hspace*{.3cm} j_{\mu, V}(x)
= \bar q_j(x) \, \gamma_\mu \, q_i(x)\,, \nonumber\\
& &J_{\mu, \, A}(q) = \int d^4x \, e^{-iqx} \, j_{\mu, \, A}(x) \,,
\quad \hspace*{.3cm} j_{\mu, A}(x) = \bar q_j(x) \, \gamma_\mu \,
\gamma_5 \, q_i(x)\,.
\en
Next, using the chiral Lagrangian derived
in section~\ref{Chir_Lagrangian}, we construct the vector/axial
currents with quantum numbers of the bare quark currents which
include mesonic degrees of freedom. Then these currents are
projected on corresponding quark (initial and final) states in order
to evaluate the dressed vector and axial quark form factors encoding 
the chiral corrections. Finally, using the dressed quark form factors 
in momentum space we can determine their Fourier--transform in
coordinate space.

The dressed quark operators $j_{\mu, \, V(A)}^{\rm dress}(x)$ 
include one-- $j_{\mu, \, V(A)}^{\rm dress (1)}(x)$ 
and two--body $j_{\mu, \, V(A)}^{\rm dress (2)}(x)$ operators: 
\eq 
j_{\mu, \, V(A)}^{\rm dress}(x) = j_{\mu, \, V(A)}^{\rm dress (1)}(x)
+ j_{\mu, \, V(A)}^{\rm dress (2)}(x) \,. 
\en  
In Ref.~\cite{Faessler:2005gd,Faessler:2006ky} we restricted to the 
consideration of one--body quark operators only. Here we discussed 
also the two--body operators. In Figs.1-4 we display the tree and 
loop diagrams which contribute to the dressed one-- and two--body 
vector and axial operators, respectively, up to and including third 
order in the chiral expansion. 

The dressed one--body quark operators $j_{\mu, \, V(A)}^{\rm dress (1)}(x)$ 
and their Fourier transforms $J_{\mu, \, V(A)}^{\rm dress (1)}(q)$ have 
the forms 
\eq\label{JVmu_dress}
j_{\mu, \, V}^{\rm dress (1)}(x) &=&
f_1^{ij}(-\partial^2) \, [ \bar q_j(x) \gamma_\mu q_i(x) ] \, + \,
\frac{f_2^{ij}(-\partial^2)}{m_i + m_j} \, \partial^\nu \, [ \bar
q_j(x) \sigma_{\mu\nu} q_i(x) ] \, - \,
\frac{f_3^{ij}(-\partial^2)}{m_i + m_j} \, i \,  \partial_\mu \,
[ \bar q_j(x) q_i(x) ] \,, \nonumber\\
J_{\mu, \, V}^{\rm dress (1)}(q) &=& \int d^4x \, e^{-iqx} \,
\bar q_j(x) \, \biggl[ \, \gamma_\mu \, f_1^{ij}(q^2)
\, + \, \frac{i \sigma_{\mu\nu} \, q^\nu}{m_i + m_j} \, f_2^{ij}(q^2) \,
\, + \, \frac{q_\mu}{m_i + m_j} \, f_3^{ij}(q^2) \, \biggr] \, q_i(x)\,,
\en
and
\eq\label{JAmu_dress}
j_{\mu, \, A}^{\rm dress (1)}(x) &=&
g_1^{ij}(-\partial^2) \, [ \bar q_i(x) \gamma_\mu \gamma_5 q_j(x) ]
\, + \, \frac{g_2^{ij}(-\partial^2)}{m_i + m_j} \, \partial^\nu \,
[ \bar q_j(x) \sigma_{\mu\nu} \gamma_5 q_i(x) ]
\, - \, \frac{g_3^{ij}(-\partial^2)}{m_i + m_j} \, i \,  \partial_\mu \,
[ \bar q_j(x) \gamma_5 q_i(x) ] \,, \nonumber\\
J_{\mu, \, A}^{\rm dress (1)}(q) &=& \int d^4x \, e^{-iqx} \,
\bar q_j(x) \, \biggl[ \, \gamma_\mu \, \gamma_5 \, g_1^{ij}(q^2)
\, + \, \frac{i \sigma_{\mu\nu} q^\nu}{m_i + m_j} \, \gamma_5
\, g_2^{ij}(q^2) \, \, + \, \frac{q_\mu}{m_i + m_j} \,
\gamma_5 \, g_3^{ij}(q^2) \, \biggr] \, q_i(x)\,,
\en
where $m_{i(j)}$ is the dressed constituent quark mass of
$i(j)$--th flavor generated by the chiral Lagrangian~(\ref{L_exp})
[see details in Ref.~\cite{Faessler:2005gd}];
$f_{1,2,3}^{ij}(q^2)$ and $g_{1,2,3}^{ij}(q^2)$ are the one-body quark
vector and axial $i \to j$ flavor changing form factors encoding the 
chiral corrections. In Figs.1 and 2 we only show the one--body diagrams 
which are relevant for the calculation of the vector 
$f_1^{ij} = f_1^{ij}(0)$
and axial $g_1^{ij} = g_1^{ij}(0)$ couplings at the order of accuracy
to which we are working in. The ellipses $\cdots$ in Figs.1 and 2 
denote higher--order diagrams, {\it i.e.}, diagrams which contribute 
only to the $q^2$--dependence of $f_1^{ij}(q^2)$ and $g_1^{ij}(q^2)$ 
and/or to the remaining four form factors $f_{2(3)}^{ij}(q^2)$ and 
$g_{2(3)}^{ij}(q^2)$. The full analysis of all six form factors goes 
beyond the scope of present paper. The contributions of the various 
graphs in Figs.1 and 2 to the vector and axial couplings is 
discussed in Appendix A and are listed in Tables 1 and 2.  
Evaluation of the one--body diagrams in Figs.1 and 2 was performed using 
the method of {\it infrared dimensional regularization} (IDR) suggested
in Ref.~\cite{Becher:1999he} in order to guarantee a straightforward
connection between loop and chiral expansions in terms of quark
masses and small external momenta. We relegate a detailed discussion
of the calculational technique to Ref.~\cite{Faessler:2005gd}. 

In Figs.3 and 4 we display the two-body diagrams (contributions to the 
vector and axial operators, respectively) which are generated by the 
chiral Lagrangian~(\ref{L_exp}) at order of accuracy we are working in. 
These diagrams include the terms generated by meson exchange and 
by the contact terms representing contributions due to heavy states and 
generated by the two-body $O(p^3)$ chiral Lagrangian 
${\cal L}_{qq}^{(3)}$~(\ref{L_exp4}). 
Inclusion of the two--body quark operators in the evaluation 
of the vector and axial couplings of baryons goes beyond the scope 
of present paper and will be done in future [therefore, in the numerical 
calculations we will restrict to the one-body approximation]. 
Here we just give the general expressions for the Fourier transforms 
of the two-body operators 
$J_{\mu, \, V(A)}^{\rm dress (2)}(q)$: 
\seq\label{J_dress_two_body}
\eq
j_{\mu, \, V}^{\rm dress (2)}(q) &=& \int d^4x \, e^{-iqx} \, 
\sum\limits_m \ ( \bar q_j(x) \, \Gamma_{ij, m}^V q_i(x) \ 
\bar q_l(x) \, \Gamma_{kl, m}^V \, q_k(x) )_\mu \ f_m^{ijkl}(q^2)\,, \\ 
j_{\mu, \, A}^{\rm dress (2)}(q) &=& \int d^4x \, e^{-iqx} \, 
\sum\limits_m \ ( \bar q_j(x) \, \Gamma_{ij, m}^A q_i(x) \ 
\bar q_l(x) \, \Gamma_{kl, m}^A \, q_k(x) )_\mu \ g_m^{ijkl}(q^2)\,, 
\en
\sen 
where $\Gamma_{ij, m}^{V(A)}$ reflects the corresponding spin structures, 
$f(g)_m^{ijkl}(q^2)$ are the two-body quark form factors encoding chiral 
effects and $m$ is the summation index over possible contributions to 
the two-body operators. We discuss the two-body operators in Appendix B. 

In order to calculate the vector and axial current transitions 
between baryons we project the dressed quark operators between the 
corresponding baryon states. The master formulas are:
\eq\label{master}
\la B_j(p^\prime)
| \, J_{\mu, \, V(A)}^{\rm dress}(q) \, | B_i(p) \ra = (2\pi)^4 \,
\delta^4(p^\prime - p - q) \, M_{\mu, \, V(A)}^{B_iB_j}(p,p^\prime)
\,, 
\en 
\eq
M_{\mu, \, V}^{B_iB_j}(p,p^\prime) &=&
\sum\limits_{k = 1}^3 f_k^{ij}(q^2) \, 
\la B_j(p^\prime)|\,V_{\mu, k}^{ij}(0) \, |B_i(p) \ra
+ \sum\limits_m f_m^{ijkl}(q^2) \, 
\la B_j(p^\prime)|\,V_{\mu, m}^{ijkl}(0) \, |B_i(p) \ra 
\nonumber\\
&=&\bar u_{B_j}(p^\prime) \biggl\{ \gamma_\mu \, F_1^{B_iB_j}(q^2)
\, + \, \frac{i \sigma_{\mu\nu} q^\nu}{m_{B_i} + m_{B_j}}
\, F_2^{B_iB_j}(q^2) + \frac{q_\mu}{m_{B_i}
+ m_{B_j}} \, F_3^{B_iB_j}(q^2) \biggr\} u_{B_i}(p) \, 
\label{M_V} \,,\\
M_{\mu, \, A}^{B_iB_j}(p,p^\prime) &=&
\sum\limits_{k = 1}^3 g_k^{ij}(q^2)
\, \la B_j(p^\prime)|\,A_{\mu, k}^{ij}(0) \, |B_i(p) \ra
+ \sum\limits_m g_m^{ijkl}(q^2) \, 
\la B_j(p^\prime)|\,A_{\mu, m}^{ijkl}(0) \, |B_i(p) \ra 
\nonumber\\
&=& \bar u_{B_j}(p^\prime) \biggl\{ \gamma_\mu \, \gamma_5 \,
G_1^{B_iB_j}(q^2) \, + \, \frac{i \, \sigma_{\mu\nu} q^\nu}{m_{B_i}
+ m_{B_j}} \, \gamma_5 \, G_2^{B_iB_j}(q^2) + \frac{q_\mu}{m_{B_i} +
m_{B_j}} \, \gamma_5 \, G_3^{B_iB_j}(q^2) \biggr\} u_{B_i}(p) 
\label{M_A} \,,
\en
where $B_i(p)$ and $u_{B_i}(p)$ are the baryon
state and spinor, respectively, normalized via 
\eq 
\la B_i(p^\prime) | B_i(p) \ra = 2 E_{B_i} \, (2\pi)^3 \,
\delta^3(\vec{p}-\vec{p}^{\,\prime})\,, \hspace*{.5cm} 
\bar u_{B_i}(p) u_{B_i}(p) = 2 m_{B_i} 
\en 
with $E_{B_i} = \sqrt{m_{B_i}^2+\vec{p}^{\,2}}$ being the baryon energy 
and $m_{B_i}$ the baryon mass. The index $i(j)$ attached to the baryon
state/field indicates the flavor of the quark involved into the
semileptonic transition. Here, $F_k^{B_iB_j}(q^2)$ and 
$G_k^{B_iB_j}(q^2)$ with $k=1,2,3$ are the vector and axial semileptonic 
form factors of baryons.

The main idea of the above relations is to express the matrix
elements of the dressed quark operators in terms of the matrix
elements of the one-- and two--body valence quark operators  
$V_{\mu, k}^{ij}(0)$, $A_{\mu, k}^{ij}(0)$, 
$V_{\mu, m}^{ijkl}(0)$ and $A_{\mu, m}^{ijkl}(0)$ and encode the chiral 
effects in the form factors  $f_k^{ij}(q^2)$, 
$g_k^{ij}(q^2)$, $f_m^{ijkl}(q^2)$ and $g_m^{ijkl}(q^2)$. 
The set of the valence quark operators is defined as 
\seq\label{Operators_VA1}
\eq
& &V_{\mu, k}^{ij}(0) = \bar q_j(0)
\Gamma^{V}_{\mu, k} q_i(0)\,, \hspace*{3.65cm} A_{\mu, k}^{ij}(0) =
\bar q_j(0) \Gamma^{A}_{\mu, k} q_i(0)\,, \\ 
& &V_{\mu, m}^{ijkl}(0) = ( \bar q_j(0) \, \Gamma_{ij, m}^V q_i(0) 
\ \bar q_l(0) \, \Gamma_{kl}^V q_k(0) )_\mu \,, \hspace*{.7cm} 
A_{\mu, m}^{ijkl}(0) = ( \bar q_j(0) \, \Gamma_{ij, m}^A q_i(0) 
\ \bar q_l(0) \, \Gamma_{kl, m}^A q_k(0) )_\mu 
\en 
\sen 
where
\eq\label{Operators_VA2} & &\Gamma^V_{\mu, 1} = \gamma_\mu\,,
\hspace*{.65cm} \Gamma^V_{\mu, 2} = \frac{i \sigma_{\mu\nu}
q^\nu}{m_i + m_j} \,, \hspace*{.65cm}
\Gamma^V_{\mu, 3} = \frac{q_\mu}{m_i + m_j} \,, \nonumber\\
& &\Gamma^A_{\mu, 1} = \gamma_\mu \gamma_5 \,, \hspace*{.3cm}
\Gamma^A_{\mu, 2} = \frac{i \sigma_{\mu\nu} q^\nu}{m_i + m_j}
\gamma_5\,, \hspace*{.3cm} \Gamma^A_{\mu, 3} =
\frac{q_\mu}{m_i + m_j} \gamma_5\,.
\en
The set of Eqs.~(\ref{master})--(\ref{Operators_VA2}) contains our main
result: we perform a separation of the effects of internal dynamics of 
valence quarks contained in the matrix elements of the bare quark 
operators $V(A)_{\mu, k}^{ij}(0)$, $V(A)_{\mu, m}^{ijkl}(0)$ and the 
effects dictated by chiral dynamics which are encoded in the relativistic
form factors $f_k^{ij}(q^2)$, $g_k^{ij}(q^2)$, $f_m^{ijkl}(q^2)$ and 
$g_m^{ijkl}(q^2)$. In particular, 
the results for the baryon properties (static characteristics and 
form factors in the Euclidean region) derived using the above
formulas satisfy the low--energy theorems and identities 
dictated by the infrared singularities of QCD [see, {\it e.g.}, 
the detailed discussion in Refs.~\cite{Faessler:2005gd,Faessler:2006ky}],   
Let us stress that consistency in the present formalism with ChPT 
is limited since we cannot consider baryonic matrix elements in Minkowski 
space. Due to the factorization of the chiral effects and effects of internal 
dynamics of valence quarks the calculation of the form factors 
$f(g)_k^{ij}(q^2)$ and $f(g)_m^{ijkl}(q^2)$ 
encoding chiral dynamics, on one side, and the matrix elements of 
$V(A)_{\mu, k}^{ij}(0)$ and $V(A)_{\mu, m}^{ijkl}(0)$ 
encoding effects of valence quarks, on the other 
side, can be done independently. The evaluation of the 
matrix elements $V(A)_{\mu, k}^{ij}(0)$ and $V(A)_{\mu, m}^{ijkl}(0)$ 
is not restricted to the small squared momenta and, therefore, can shed 
light on baryon form factors at higher Euclidean 
momentum squared in comparison with ChPT. In particular, as a first step 
we employ a formalism motivated by the ChPT Lagrangian,  
which is formulated in terms of constituent quark degrees of freedom,  
for the calculation of $f(g)_k^{ij}(q^2)$ and $f(g)_m^{ijkl}(q^2)$.  
The calculation of the matrix elements of the bare quark operators can  
then be relegated to quark models based on specific assumptions about  
internal quark dynamics, hadronization and confinement. Note that  
Eqs.~(\ref{master})--(\ref{Operators_VA2}) are valid for the calculations  
of dressed vector and axial quark operators of {\it any} flavor content. 

\subsection{Chiral expansion of vector and axial quark couplings}

In this section we present the results for the chiral expansion of
the vector $f_1^{ij} = f_1^{ij}(0)$ and axial $g_1^{ij} = g_1^{ij}(0)$ 
quark couplings for various $i \to j$ flavor transitions in 
the SU(2) (isospin) limit. We begin by defining the quark wave function 
renormalization constant $Z$. In particular, the tree graphs 
$T_{\rm tree}$ in Figs.1(a) and 2(a) should be renormalized in terms 
of $Z$ via $1/2 \{ T_{\rm tree}, Z\}$. Details of the calculation of $Z$ 
can be found in~\cite{Faessler:2005gd}. In SU(2) we find
\eq
Z = 1 - \frac{9}{4}
R_\pi \en while in SU(3) we have \eq {\rm diag}\{Z, Z, Z_s\} = I -
\sum\limits_{P = \pi, K, \eta} \alpha_P R_P
\en
where in both cases $Z \equiv Z_u \equiv Z_d$ and
\eq\label{R_P}
& &R_P =
\frac{g^2}{F^2} \Delta_P + \frac{g^2 M_P^2}{24 \pi^2 F^2} \biggl\{ 1
- \frac{3 \pi}{2} \mu_P \biggr\}\,, \hspace*{.5cm} \mu_P =
\frac{M_P}{m}\,,
\hspace*{.5cm} P = \pi, K, \eta \,, \nonumber\\
& &\alpha_\pi = \frac{9}{2} Q + \frac{3}{2} I - \frac{9}{4}
\lambda_3\,, \hspace*{.5cm} \alpha_K   = - 3 Q + 2 I + \frac{3}{2}
\lambda_3\,, \hspace*{.5cm} \alpha_\eta= - \frac{3}{2} Q +
\frac{1}{2} I + \frac{3}{4} \lambda_3 \,.
\en  
Here $Q = {\rm diag}\{2/3, -1/3, -1/3\}$ and
the quantity $\Delta_P$ is defined as
\eq
\Delta_P = 2 M_P^2 \, \lambda_{P}\,, \hspace*{.5cm} \lambda_{P}
&=& \frac{M_P^{d-4}}{16 \pi^2} \, \biggl\{ \frac{1}{d-4} -
\frac{1}{2} (\ln 4\pi + \Gamma^\prime(1) + 1)\biggr \} \,.
\en
In Appendix A we explicitly list the contributions of the various
graphs to the vector (Fig. 1) and axial (Fig. 2) couplings. 
First, we discuss the vector couplings. For completeness, we consider 
vector currents conserving the quark flavor, 
{\it i.e.}, corresponding to electric and isospin charge, 
as well as those involving $d \to u$ and $s \to u$
transitions.  Due to charge conservation and isospin invariance, the
total contribution of the diagrams of Fig.1 properly reproduces the
quark electric and isospin charges.  In addition, the vector
coupling governing the $d \to u$ transition is equal to
unity --- $f_1^{du} = 1$ [a detailed discussion for both SU(2) and
SU(3) is presented in Appendix A].  As stressed in
Ref.~\cite{Villadoro:2006nj} the total contribution of diagrams
shown in Fig.1 to these quantities is finite and no unknown LEC's
appear at the order of accuracy to which we are working in. In the case
of the $s \to u$ transition, the total contribution of the diagrams
given in Fig.1 to the corresponding vector coupling $f_1^{su}$ is
finite but contains symmetry breaking corrections of second order in
SU(3) --- ${\cal O}((M_K-M_\pi)^2)$ and ${\cal O}((M_K-M_\eta)^2)$.
Note, that the Ademollo--Gatto theorem (AGT) protects the coupling 
$f_1^{su}$ from {\it first--order} symmetry breaking corrections. 
As shown explicitly [{\it cf.} Appendix A] the AGT holds for the two
sets of diagrams set I and set II independently.  For set I,
including the diagrams of Fig.1(a), (b), (e), and (f) and for II,
including the diagrams of Fig.1(c) and (d) we have: 
\eq
f_1^{su; I} = \sum\limits_{i = a, b, e, f} f_1^{su; (i)} =
1 - \frac{9 g^2}{16} ( H_{\pi K} + H_{\eta K} + G_{\pi K} + G_{\eta K} )
\en
and
\eq
f_1^{su; II} = \sum\limits_{i = c, d} f_1^{su; (i)} = - \frac{3}{16}
( H_{\pi K} + H_{\eta K} ) \,. 
 \en 
The ${\cal O}(p^2)$ functions $H_{ab}$ and $G_{ab}$,
which show up in the context of ChPT [see, {\it e.g.},
Refs.~\cite{Leutwyler:1984je,Krause:1990xc,Anderson:1993as,%
Kaiser:2001yc,Villadoro:2006nj,Lacour:2007wm,Gasser:1984ux}], 
are defined as 
\eq\label{fun_HG}
H_{ab} &=& \frac{1}{(4 \pi F)^2}
\biggl( M_a^2 + M_b^2 - \frac{2 M_a^2 M_b^2}{M_a^2 - M_b^2}
{\rm ln}\frac{M_a^2}{M_b^2} \biggr)
= {\cal O}( (M_a^2 - M_b^2)^2 )\,, \nonumber\\
G_{ab} &=& - \frac{1}{(4 \pi F)^2} \frac{2\pi}{3m}
\frac{(M_a - M_b)^2}{M_a + M_b} (M_a^2 + 3 M_a M_b + M_b^2) =
{\cal O}( (M_a^2 - M_b^2)^2 ) \,.
\en
Therefore, the final result for the $s \to u$
quark transition vector coupling is
\eq\label{f1_delta}
f_1^{su} =
f_1^{su; I} + f_1^{su; II} =  1 + \delta f_1^{su} = 1 - \frac{3}{16}
\biggl( (1 + 3 g^2) ( H_{\pi K} + H_{\eta K} ) + 3 g^2 ( G_{\pi K} +
G_{\eta K} ) \biggr)
\en
where $\delta f_1^{su}$ is the total $SU(3)$ breaking correction.

Next we turn to the discussion of the axial couplings governing the $d
\to u$ and $s \to u$ quark flavor transitions.  The expressions for
the axial (isovector) charge $g_1$ and the axial couplings
responsible for the $d \to u$ and $s \to u$ transitions $g_1^{du}$ 
and $g_1^{su}$ are given in the following [{\it cf.} Appendix A for 
the expressions of the separate diagrams in Fig.2].  In SU(2) we have 
\eq 
g_1 = g_1^{du} = g \biggl\{ 1 - \frac{g^2 M_\pi^2}{16
\pi^2 F^2} + \frac{M_\pi^3}{24 \pi m F^2} \biggl( 3 + 3 g^2 - 4
c_3^q m + 8 c_4^q m \biggr) \biggr\} + 4 M_\pi^2 \biggl\{ d_{16}^q -
\frac{g}{F^2} \biggr( \frac{1}{2} + g^2 \biggr) \lambda_\pi \biggr\}
\,.
\en
Absorbing the infinity in the LEC $d_{16}^q$
\eq
d_{16}^q = \bar d_{16}^q + \frac{g}{F^2} \biggr( \frac{1}{2} + g^2
\biggr) \lambda_\pi
\en
we arrive at the ultraviolet-finite (UV) expression for $g_1$:
\eq
g_1 = g_1^{du} = g + \delta g_1^{\chi_{_2}} = g
\biggl\{ 1 - \frac{g^2 M_\pi^2}{16 \pi^2 F^2} + \frac{M_\pi^3}{24
\pi m F^2} \biggl( 3 + 3 g^2 - 4 c_3^q m + 8 c_4^q m \biggr)
\biggr\} + 4 M_\pi^2 \bar d_{16}^q \,, 
\en
where $\delta g_1^{\chi_{_2}}$ is the SU(2) chiral correction.

In SU(3) the corresponding expression for the isovector axial
coupling $g_1$ is:
\eq\label{g1_du}
g_1 &=& g_1^{du} = g
\biggl\{ 1 - \frac{g^2}{16 \pi^2 F^2}
\biggl(M_\pi^2 + M_K^2
+ \frac{M_\eta^2}{3} \biggr) \nonumber\\
&+& \frac{M_\pi^3}{24 \pi m F^2}
\biggl( 3 + 3 g^2 - 4 C_3^q m + 8 C_4^q m \biggr)
+ \frac{M_K^3}{48 \pi m F^2}
\biggl( 3 + \frac{9}{2} g^2 + 8 C_4^q m \biggr)
+ \frac{g^2 M_\eta^3}{48 \pi m F^2} \biggr\} \nonumber\\
&+& 2 M_\pi^2 \biggl( D_{16}^q + \frac{1}{3} D_{17}^q
- \frac{g}{F^2} \biggl( 1 + \frac{17}{9} g^2 \biggr) \bar\lambda
- \frac{g ( 1 + 2 g^2 )}{32 \pi^2 F^2} \ln\frac{M_\pi^2}{m^2}
+ \frac{g^3}{288 \pi^2 F^2} \ln\frac{M_\eta^2}{m^2} \biggr) \nonumber\\
&+& 2 M_K^2 \biggl( 2 D_{16}^q - \frac{1}{3} D_{17}^q -
\frac{g}{2F^2} \biggl( 1 + \frac{35}{9} g^2 \biggr) \bar\lambda -
\frac{g ( 1 + 3 g^2 )}{64 \pi^2 F^2} \ln\frac{M_K^2}{m^2} -
\frac{g^3}{72 \pi^2 F^2} \ln\frac{M_\eta^2}{m^2} \biggr) \,,
\en
where the divergent quantity
\eq
\bar\lambda &=& \frac{m^{d-4}}{16\pi^2} \,
\biggl\{ \frac{1}{d-4} - \frac{1}{2} (\ln 4\pi +
\Gamma^\prime(1) + 1)\biggr \} \,, 
\en
coincides with $\lambda_P$ when $m = M_P$.
The last two lines in Eq.(\ref{g1_du}), containing
divergences, can be written in more compact form in terms of one of
the divergent quantities $\lambda_P$ ($P = \pi, K, \eta$) --- {\it e.g.}, 
in terms of $\lambda_\eta$, these lines have the succinct form
\eq
& & 2 M_\pi^2 \biggl( D_{16}^q + \frac{1}{3} D_{17}^q -
\frac{g}{F^2} \biggl( 1 + \frac{17}{9} g^2 \biggr)
\lambda_\eta - (1 + 2 g^2) L_{\pi\eta} \biggr) \nonumber\\
&+& 2 M_K^2 \biggl( 2 D_{16}^q  - \frac{1}{3} D_{17}^q -
\frac{g}{2F^2} \biggl( 1 + \frac{35}{9} g^2 \biggr) \lambda_\eta -
\frac{1}{2} (1 + 3 g^2) L_{K\eta} \biggr) \,, 
\en 
where 
\eq
L_{ab} = \frac{g}{32 \pi^2 F^2} \ln\frac{M_a^2}{M_b^2} 
\en 
Again we remove the divergences in $g_1$ using the renormalized LEC's
$\bar D_{16}^q$ and $\bar D_{17}^q$: \eq D_{16}^q &=& \bar D_{16}^q
+ \frac{g}{2 F^2}
\biggl( 1 + \frac{23}{9} g^2 \biggr) \lambda_\eta \,, \nonumber\\
D_{17}^q &=& \bar D_{17}^q + \frac{3 g}{2 F^2} \biggl( 1 +
\frac{11}{9} g^2 \biggr) \lambda_\eta \,.
\en
Therefore, the result
for the renormalized coupling $g_1^{du}$ in SU(3) is expressed in
terms of the axial charge and quark mass in the chiral limit, meson
masses and two unknown LEC's $\bar D_{16}$ and $\bar D_{17}$:
\eq
g_1 &=& g_1^{du} = g + \delta g_1^{\chi_{_3}} =
g \biggl\{ 1 - \frac{g^2}{16 \pi^2 F^2}
\biggl(M_\pi^2 + M_K^2
+ \frac{M_\eta^2}{3} \biggr) \nonumber\\
&+& \frac{M_\pi^3}{24 \pi m F^2}
\biggl( 3 + 3 g^2 - 4 C_3^q m + 8 C_4^q m \biggr)
+ \frac{M_K^3}{48 \pi m F^2}
\biggl( 3 + \frac{9}{2} g^2 + 8 C_4^q m \biggr)
+ \frac{g^2 M_\eta^3}{48 \pi m F^2} \biggr\} \nonumber\\
&+& 2 M_\pi^2 \biggl( \bar D_{16}^q + \frac{1}{3} \bar D_{17}^q - (1
+ 2 g^2) L_{\pi\eta} \biggr) + 2 M_K^2 \biggl( 2 \bar D_{16}^q  -
\frac{1}{3} \bar D_{17}^q - \frac{1}{2} (1 + 3 g^2) L_{K\eta}\biggr)
\en
where $\delta g_1^{\chi_{_3}}$ is the SU(3) chiral correction.

Also we perform an expansion of $g_1 = g_1^{du}$ in the powers of
the SU(3) breaking parameter $m_s - \hat m$:
\eq
g_1 = g_1^{du} = g_1^{\rm SU_3} + \delta g_1
\en
where
\eq
g_1^{\rm SU_3} = g \biggl\{ 1 - \frac{7 g^2 \bar M^2}{48 \pi^2 F^2}
+ \frac{\bar M^3}{48 \pi m F^2}
\biggl( 9 + \frac{23}{2} g^2 - 8 C_3^q m + 24 C_4^q m \biggr) \biggr\}
+ 6 \bar M^2 \bar D_{16}^q
\en
is the SU(3) symmetric term, and
\eq\label{delta_g1du}
\delta g_1 &=& (M_K^2 - M_\pi^2)
\biggl\{ \frac{g}{96 \pi^2 F^2} \biggl( 9 + \frac{59}{3} g^2 \biggr)
- \frac{g \bar M}{96 \pi m F^2}
\biggl( 9 + \frac{11}{2} g^2 - 16 C_3^q m + 24 C_4^q m \biggr)
- \frac{2}{3} \bar D_{17}^q \biggr\} \nonumber\\
&+& {\cal O}( (M_K^2 - M_\pi^2)^2 ) \equiv h_1 (M_K^2 - M_\pi^2) +
{\cal O}( (M_K^2 - M_\pi^2)^2 ) \,, 
\en 
is the SU(3) breaking term,
where we display the first--order term. Here for convenience 
we define the so--called SU(3) symmetric octet mass $\bar M$ of
pseudoscalar mesons as $\bar M^2 = 2 \bar m B$ with $\bar m = (m_u +
m_d + m_s)/3 = (2 \hat m + m_s)/3$.

Likewise, we have the result for the $s \to u$ flavor transition
axial coupling $g_1^{su}$ in terms of $\bar D_{16}^q$ and
$\bar D_{17}^q$:
\eq
g_1^{su} &=& g + \delta g_1^{\chi_{su}} =
g \biggl\{ 1 - \frac{3 g^2}{64 \pi^2 F^2}
\biggl(M_\pi^2 + 2 M_K^2
+ \frac{M_\eta^2}{9} \biggr)
+ \frac{M_\pi^3}{64 \pi m F^2}
\biggl( 3 + \frac{9}{2} g^2 + 16 C_4^q m \biggr) \nonumber\\
&+& \frac{M_K^3}{32 \pi m F^2}
\biggl( 3 + \frac{9}{2} g^2 - \frac{16}{3} C_3^q m
+ \frac{16}{3} C_4^q m \biggr) + \frac{M_\eta^3}{64 \pi m F^2}
\biggl( 3 + \frac{11}{6} g^2 + \frac{16}{3} C_4^q m \biggr) \biggr\}
\nonumber\\
&+& 2 M_\pi^2 \biggl( \bar D_{16}^q - \frac{1}{6} \bar D_{17}^q -
\frac{3}{8} (1 + 3 g^2) L_{\pi\eta} \biggr) + 2 M_K^2 \biggl( 2 \bar
D_{16}^q  + \frac{1}{6} \bar D_{17}^q - \frac{3}{4} (1 + 3 g^2)
L_{K\eta} \biggr) \,,
\en
where $\delta g_1^{\chi_{su}}$ is the strangeness
changing chiral correction. The $m_s - \hat m$ expansion for
the $g_1^{su}$ coupling reads:
\eq
g_1^{su} = g_1^{\rm SU_3} + \delta g_1^{su}
\en
where
\eq\label{delta_g1su}
\delta g_1^{su} &=& (M_K^2 - M_\pi^2)
\biggl\{ \frac{g}{192 \pi^2 F^2} \biggl( 9 + \frac{79}{3} g^2 \biggr)
+ \frac{g \bar M}{192 \pi m F^2}
\biggl( 9 + \frac{11}{2} g^2 - 16 C_3^q m - 16 C_4^q m \biggr)
+ \frac{1}{3} \bar D_{17}^q \biggr\} \nonumber\\
&+& {\cal O}( (M_K^2 - M_\pi^2)^2 ) \equiv h_2 (M_K^2 - M_\pi^2) +
{\cal O}( (M_K^2 - M_\pi^2)^2 ) \,. \en

\subsection{Bare quark matrix elements}

Now we are in the position to discuss the calculation of the matrix 
elements of the bare quark operators derived in Eqs.~(\ref{M_V}) 
and (\ref{M_A}) and restrict in the following to the one-body approximation: 
\eq\label{Operators_V1A1} 
V_{\mu, 1}^{ij}(0) = \bar q_j(0) \gamma_\mu q_i(0)\,, 
\hspace*{1cm} A_{\mu, 1}^{ij}(0) = \bar q_j(0) 
\gamma_\mu \gamma_5 q_i(0)\,.  
\en  
As stressed earlier, the evaluation of these matrix elements can be  
done independently of the calculation of chiral effects. Therefore,   
it can be relegated to a quark model based on a specific scenario   
about hadronization and confinement of quarks within baryons  
including internal quark dynamics. 

As mentioned in the Introduction, in the calculation of the baryon
matrix elements $\la \,V_{\mu, 1} \, \ra = \la B_j(p^\prime)|
\,V_{\mu, 1}^{ij}(0) \, |B_i(p) \ra$ and $\la \,A_{\mu, 1} \, \ra =
\la B_j(p^\prime)|\,A_{\mu, 1}^{ij}(0) \, |B_i(p) \ra$ we employ a
general ansatz for the spatial form of the quark wave functions, 
without referring to any specific realization.  In a forthcoming
paper~\cite{Faessler:2007qm} we will evaluate the baryon matrix
elements using a Lorentz-- and gauge--invariant quark model
based on a specific hadronization ansatz -- i.e. modeling internal 
quark dynamics, which goes beyond additive quark model. 
Note that in Ref.~\cite{Faessler:2006ky} we did an analogous study 
of the electromagnetic properties of the baryon octet and the 
$\Delta(1230)$--resonance. One should stress that 
the approach~\cite{Faessler:2006ky} is resticted to the evaluation 
of baryon matrix elements in the Euclidean space to avoid unphysical 
cuts. Therefore, it pretends to the evaluation of the baryon 
matrix elements only for the Euclidean transfered momentum squared. 

In the evaluation of the bare matrix elements $\la \,V_{\mu, 1} \,
\ra$ and $\la \,A_{\mu, 1} \, \ra$ we follow, {\it e.g.},
Refs.~\cite{Donoghue:1992dd,PCQM}.   We begin by introducing the
ground--state wave function of the quark with flavor $f$ moving in a
spin--independent central potential:
\eq\label{eigenstate} 
q_f(x) \, = \, q_f(\vec{x}) \, e^{-iEt}\,, 
\hspace*{1cm} q_f(\vec{x}) \, = \,
\, \left(
\begin{array}{c}
u_f(r)\\
i l_f(r) \, \frac{\vec{\sigma}\cdot\vec{x}}{r}\\
\end{array}
\right) \, \chi_s \, \chi_f \, \chi_c \,,
\en
where $u_f(r)$ and $l_f(r)$ signify the upper and lower components 
of the quark wave function (in the nonrelativistic limit $l_f$ 
vanishes); $\chi_s$, $\chi_f$, $\chi_c$ are the spin, flavor, and 
color quark wave functions, respectively.  Note that this form of 
the quark wave function also appears in relativistic harmonic 
oscillator models 
utilizing a central potential [see references in~\cite{PCQM}]. 
In the following we use the notation  $(u = u_u = u_d, \, l = l_u =
l_d)$ for nonstrange and $(u_s, \, l_s)$ for strange quark
wave functions. In practice it is also convenient to introduce ratios
between the two sets of wave functions via: 
\eq\label{ratio_ul} 
\xi_u = \frac{u_s}{u}\,, \hspace*{1cm} \xi_l = \frac{l_s}{l}\,. 
\en 
The normalization condition for the spatial wave function is:
\eq\label{quark_norm}
\int d^3 x \, q^\dagger_f(\vec{x})
q_f(\vec{x}) = \int d^3 x \, ( u^2_f(r) + l^2_f(r) ) = 1 \,.
\en
Now we are in the position to pin down the matrix elements
$\la \, V_{\mu, 1} \, \ra$ and $\la \, A_{\mu, 1} \, \ra$ by considering
quark operators with different flavor structure.  We begin by calculating
the vector matrix elements for the respective initial and final baryon 
states:
\eq\label{V1BiBj}
V_1^{B_iB_j} &=& \la B_j \uparrow | \int
d^3 x \,
\bar q_j(x) \gamma^0 q_i(x) | B_i \uparrow \ra \, \nonumber\\
&=& \int d^3 x \biggl[ u_i(r) u_j(r) + l_i(r) l_j(r) \biggr] \, \la
B_j \uparrow | \sum\limits_{k=1}^3 \, \lambda_{ji}^k | B_i \,
\uparrow \ra \,, 
\en 
where the spin--flavor matrix elements $\la B_j
\uparrow | \sum\limits_{k=1}^3 \, \lambda_{ji}^k | B_i \uparrow \ra$
are evaluated using the simple SU(6) quark model: 
\eq 
\la B_j \uparrow | \sum\limits_{k=1}^3 \,  
\lambda_{ji}^k | B_i  \uparrow \ra
= \, \left\{
\begin{array}{l l}
1 & {\rm for} \ n \to p \\
- \sqrt{3/2} & {\rm for} \ \Lambda \to p \\
- 1 & {\rm for} \ \Sigma^- \to n \\
0 & {\rm for} \ \Sigma^- \to \Lambda \\
  \sqrt{3/2} & {\rm for} \ \Xi^- \to \Lambda \\
 \sqrt{1/2} & {\rm for} \ \Xi^- \to \Sigma^0 \\
1 & {\rm for} \ \Xi^0 \to \Sigma^+ ,\\
\end{array}
\right .
\en
$\lambda^k_{ji}$ are the linear combinations of the Gell-Mann flavor 
matrices, and relativistic effects are included in the overlap of the 
spatial quark wave functions.  It is clear that $F_1^{np}(0) = 1$ as 
required by conservation of the vector current (CVC) --- the CVC 
prediction of unity emerges as the normalization 
condition (\ref{quark_norm}) for the spatial quark wave functions.  
Note that in the SU(3) limit, wherein quarks, independent of their 
flavor, have identical wave functions, the ``spatial'' integral 
in Eq.~(\ref{V1BiBj}) is identical to the wave function normalization 
condition (\ref{quark_norm}). As a result in $\Delta S = 1$ transitions 
the corresponding vector current is also conserved to the extent that 
the $s$-- and $u$--quark are degenerate in mass.

In the case of the corresponding baryon axial matrix elements, 
we have 
\eq
A_1^{B_iB_j} &=& \la B_j \uparrow | \int d^3 x \, \bar q_j(x)  
\gamma^3 \gamma^5 q_i(x) | B_i \uparrow \ra \, \nonumber\\ 
&=& \int d^3 x \, \biggl[ u_i(r) u_j(r) - \frac{1}{3}l_i(r) l_j(r) 
\biggr] \, \la B_j \uparrow | \sum\limits_{k=1}^3 \, \sigma_3^k 
\lambda_{ji}^k | B_i \uparrow  \ra \,,  
\en 
where the spin--flavor matrix elements 
$\la B_j \uparrow | \sum\limits_{k=1}^3 \,
\sigma_3^k  \lambda_{ji}^k | B_i  \uparrow \ra$ 
are evaluated using SU(6): 
\eq 
\la B_j \uparrow | \sum\limits_{k=1}^3 \, \sigma_3^k
\lambda_{ji}^k | B_i \uparrow \ra = \, \left\{
\begin{array}{l l}
5/3 & {\rm for} \ n \to p \\
- \sqrt{3/2} & {\rm for} \ \Lambda \to p \\
1/3 & {\rm for} \ \Sigma^- \to n \\
 \sqrt{2/3}  & {\rm for} \ \Sigma^- \to \Lambda \\
 \sqrt{1/6} & {\rm for} \ \Xi^- \to \Lambda \\
 5/(3\sqrt{2}) & {\rm for} \ \Xi^- \to \Sigma^0 \\
 5/3 & {\rm for} \ \Xi^0 \to \Sigma^+ \\
\end{array}
\right .
\en
and, again, relativistic effects are included in the
overlap of the spatial quark wave functions.

Using the normalization condition the integrals over the spatial quark 
wave functions can be simplified.  There are three possible situations:
1) overlap of nonstrange quark wave functions; 2) overlap of strange
quark wave functions; 3) overlap of nonstrange with strange quark
wave functions.  In the first case we have
\eq\label{I_VA} 
I_V &=&
\int d^3 x \biggl[ u^2(r) + l^2(r) \biggr] = 1\,
\label{IV} \\
I_A &=& \int d^3 x \biggl[ u^2(r) - \frac{1}{3} l^2(r) \biggr] = 1 -
\frac{4}{3} \int d^3 x \, l^2(r)  \,. \label{IA}
\en
In the second case:
\eq\label{I_VA_ss} 
I_V^{ss} &=& \int d^3 x \biggl[ u_s^2(r) +
l_s^2(r) \biggr] = \int d^3 x \biggl[ \xi^2_u u^2(r) + \xi^2_l
l^2(r) \biggr] = 1 \,, \label{IVss} \\
I_A^{ss} &=& \int d^3 x \biggl[ u^2_s(r) - \frac{1}{3} l^2_s(r)
\biggr] = 1 - \frac{4}{3} \xi^2_l \int d^3 x \, l^2(r) = 1 + \xi^2_l
( I_A - 1 ) \,. \label{IAss}
\en
where $\xi_u$ and $\xi_l$ can be
written in terms of the axial structure integral $I_A$ via:
\eq\label{xi_u} 
\xi_u = \sqrt{1 + 3 (1 -\xi_l^2) \frac{1 - I_A}{1 +3I_A}}  \,. 
\en 
It is clear that in the SU(3) limit --- $\xi_u = \xi_l \to1$ 
--- the expressions for the overlap integrals $I_V^{ss}$ and $I_A^{ss}$ 
reduce to $I_V$ and $I_A$, respectively. 

In the third case we have:
\eq\label{I_VA_s}
I_V^{s} &=& \int d^3 x \biggl[ u(r) u_s(r) + l(r) l_s(r) \biggr]
= \int d^3 x \biggl[ \xi_u u^2(r) + \xi_l l^2(r) \biggr]
= \xi_u + \frac{3}{4} \, (\xi_u - \xi_l) \, ( I_A - 1 ) \,,
\label{IVs} \\
I_A^{s} &=& \int d^3 x \biggl[ u(r) u_s(r) - \frac{1}{3} l(r) l_s(r)
\biggr] = \xi_u - \biggl(\xi_u + \frac{\xi_l}{3}\biggr) \int d^3 x
\, l^2(r) = \xi_u + \biggl( \frac{3\xi_u}{4} + \frac{\xi_l}{4}
\biggr) ( I_A - 1 ) \,. \label{IAs}
\en
Here the parameter $\xi_u$
can be rewritten by using identity~(\ref{xi_u}). Therefore, all
structure integrals ($I_V$, $I_A$, $I_V^s$, $I_A^s$, $I_V^{ss}$,
$I_A^{ss}$), involving spatial quark wave functions are either fixed
precisely (like $I_V = I_V^{ss} = 1$) or are expressed in terms of
$I_A$ and the parameter $\xi_l$. In the case of
exact $SU(3)$ symmetry the vector and axial integrals are
degenerate --- $I_V = I_V^s = I_V^{ss} = 1$ and $I_A = I_A^s =
I_A^{ss}$.  In the nonrelativistic limit $I_A = 1$, $\xi_u = 1$ and
all these structure integrals are unity ---
\eq
I_V = I_V^s = I_V^{ss} = I_A = I_A^s = I_A^{ss} = 1 \,.
\en
It should be stressed that the all
vector integrals satisfy the AGT --- either they are exactly equal to
unity (like $I_V$ and $I_V^{ss}$), or deviate from 1 by the
corrections of second-order in SU(3) breaking. Specifically, 
\eq
I_V^s = 1 + \frac{3}{2} \frac{I_A - 1}{1 + 3 I_A} \delta^2
+ {\cal O}(\delta^3) \,, 
\en
where $\delta = \xi_l - 1$ is a SU(3) breaking parameter.
In the case of the axial overlap integrals $I_A^s$ and $I_A^{ss}$
the SU(3) breaking corrections begin at order ${\cal O}(\delta)$. 

Finally we note that the bare matrix elements (which contain the
effects of valence quarks) can be expressed in terms of the axial
structure integral $I_A$ and the parameter $\delta$, which encode the
effects of SU(3) breaking, {\it i.e.}, distinguish the lower
components of the strange and nonstrange quark ---
\eq
I_V = I_V^{ss} = 1 \,, \hspace*{.75cm}
I_V^s = 1 + \delta I_V^s \,, \hspace*{.75cm}
I_A^s = I_A + \delta I_A^s \,, \hspace*{.75cm}
I_A^{ss} = I_A + \delta I_A^{ss}\,,
\en
where
\eq\label{delta_IV_IA}
\delta I_V^s &=& \xi_u -1 + \frac{3}{4} \, (\xi_u - \xi_l) \,
( I_A - 1 ) =  \frac{3}{2} \frac{I_A - 1}{1 + 3 I_A} \delta^2
+ {\cal O}(\delta^3) = {\cal O}(\delta^2) \,,
\nonumber\\[1mm]
\delta I_A^s &=& \xi_u - 1 + (1 - I_A)
\biggl( 1 - \frac{3\xi_u}{4} - \frac{\xi_l}{4} \biggr)
= (I_A - 1) \delta + {\cal O}(\delta^2)
= {\cal O}(\delta) \,, \\[1mm]
\delta I_A^{ss} &=& (1 - I_A) ( 1 - \xi_l^2 )
= 2 (I_A - 1) \delta + {\cal O}(\delta^2)
= {\cal O}(\delta) \,.
\nonumber
\en
In Sec.~III [see Eqs.~(\ref{match_1}), (\ref{IA_matchning})
and (\ref{delta_constraint})] doing the matching of our
results to the model--independent expressions derived in ChPT and
by Ademollo and Gatto in Ref.~\cite{Ademollo:1964sr} we express
the quantities $I_A$ and $\delta = \xi_l - 1$ in terms of parameters
of the chiral Lagrangian~(\ref{L_qU}).

\section{Semileptonic vector and axial couplings of baryons}

In this section we combine chiral and valence quarks effects in
order to derive the expressions for vector $g_V^{B_iB_j}$ and
axial $g_A^{B_iB_j}$ couplings which govern the semileptonic
transitions between octet baryons. 
Note, we use the phase convention~\cite{Leutwyler:1984je} which gives 
e.g. the positive sign for the axial coupling $g_A^{np}$ of the 
neutron $\beta$-decay. In particular, neglecting contributions of order 
$q = p^\prime - p$ the matrix elements of semileptonic decays of the 
baryon octet is determined by two constants $g_V^{B_iB_j}$ and 
$g_A^{B_iB_j}$ as: 
\eq 
M_{\mu, \, V-A}^{B_iB_j}(p,p) = 
M_{\mu, \, V}^{B_iB_j}(p,p) - M_{\mu, \, A}^{B_iB_j}(p,p) 
= \bar u_{B_j}(p) \gamma_\mu  ( g_V^{B_iB_j}  
\, - \, \gamma_5 \, g_A^{B_iB_j} ) u_{B_i}(p) \label{phase_conv}\,. 
\en 
Using Eqs.~(\ref{M_V}), (\ref{M_A}) and the expressions for the 
couplings encoding chiral effects and valence quark contributions, 
the quantities $g_V^{B_iB_j}$ and $g_A^{B_iB_j}$ are defined as:
\eq\label{gV_gA}
g_V^{B_iB_j} =
F_1^{B_iB_j}(0) = f_1^{ij} \, V_1^{B_iB_j} \,, \hspace*{.5cm}
g_A^{B_iB_j} = G_1^{B_iB_j}(0) = g_1^{ij} \, A_1^{B_iB_j} \,.
\en 

\subsection{Nucleon axial charge}

First we examine the nucleon axial charge and perform the matching
to ChPT --- we relate the parameters of our Lagrangian to those of
ChPT.  In SU(2) the expression for the nucleon axial charge is
\eq\label{ga_ChPT}
g_A = \gaz - \frac{\gaz^{\!\!\! 3} M_\pi^2}{16 \pi^2 F^2} 
+ \frac{\gaz M_\pi^3}{24 \pi \!\mnz \!\! F^2} 
\biggl( 3 + 3 \gaz^{\!\!\! 2} - 4 c_3 \mnz + 8 c_4 \mnz \biggr) 
+ 4 M_\pi^2 \bar d_{16}  \,
\en
in ChPT~\cite{Kambor:1998pi}--\cite{Becher:2001hv},
and
\eq\label{ga_our}
g_A = g_1 A_1^{np} = \frac{5}{3}
I_A \biggl\{ g  - \frac{g^3 M_\pi^2}{16 \pi^2 F^2} 
+ \frac{g M_\pi^3}{24 \pi m F^2} 
\biggl( 3 + 3 g^2 - 4 c_3^q m + 8 c_4^q  m \biggr) 
+ 4 M_\pi^2 \bar d_{16}^q \biggr\} \,
\en
in our approach, 
where $\gaz$ and $\mnz$ are the values of the nucleon axial charge
and nucleon mass in the chiral limit.  Matching these expressions
for the axial charge up to order ${\cal O}(p^4)$, we derive the 
following relations between the parameters of the two approaches:
\eq\label{matching_ga}
& &\gaz = g R =
\frac{5}{3} \, g \, I_A = \frac{5}{3} \, g \, 
\biggl( 1 - \frac{4}{3} \int d^3 x \, l^2(r) \biggr) \,,
\label{match_1}\\
& &\bar d_{16} - \frac{\gaz^{\!\!\! 3}}{64 \pi^2 F^2} =
   \frac{5}{3} I_A \biggl(\bar d_{16}^q
- \frac{g^3}{64 \pi^2 F^2} \biggr) \,, \label{match_2}\\
& & \frac{1 + \gaz^{\!\!\! 2}}{8 \!\mnz} + \frac{c_4}{3} -
\frac{c_3}{6} = \frac{1 + g^2}{8m} + \frac{c_4^q}{3} -
\frac{c_3^q}{6} \,. \label{match_3}
\en
Here for convenience we
introduce the definition $R = \gaz/g$.  Then from the first matching
condition we can derive constraints on the ``axial'' integral $I_A$
(\ref{IA}) and on the integral over the square of the lower/upper
components of the spatial quark wave functions ---
\eq\label{IA_matchning}
I_A = \frac{3}{5}R \en and
\eq\label{spatial} \int d^3 x \, l^2(r) =
1 - \int d^3 x \, u^2(r) = \frac{3}{4} \, \biggl( 1 - \frac{3}{5} R
\biggr) \,.
\en
Therefore, the second matching condition
(\ref{match_2}) reduces to
\eq\label{matching_d16}
\bar d_{16} - \frac{\gaz^{\!\!\! 3}}{64 \pi^2 F^2} = 
R \biggl( \bar d_{16}^q - \frac{g^3}{64 \pi^2 F^2} \biggr) \,.
\en
Taking into account the
relation between the mass and axial charge both of the nucleon and the 
quark in the chiral limit  
\eq\label{m_mN}
\frac{\mnz}{m} = \biggl(\frac{\gaz}{g}\biggr)^2 = R^2
\en 
as derived in Ref.~\cite{Faessler:2005gd} from 
the matching of the nucleon mass in the two approaches, 
the third condition (\ref{match_3}) can be simplified as
\eq\label{matching_c3c4}
c_3 - 2 c_4 = c_3^q - 2 c_4^q +
\frac{3}{4 \mnz} ( 1 - R^2 ) \,.
\en
We have two essential remarks:
1) the matching condition (\ref{m_mN}) is very important in our approach,
because it allows us to remove the unknown scale
parameter --- constituent quark mass --- from the explicit expressions of 
the matrix elements;
2) for the evaluation of the nucleon axial charge we do not require an
explicit form for the spatial quark wave functions [see
Eq.~(\ref{spatial})].

Having dealt with SU(2), we note the corresponding expression for
the nucleon axial charge in SU(3):
\eq
g_A &=& g R \biggl\{ 1 - \frac{g^2}{16 \pi^2 F^2}
\biggl(M_\pi^2 + M_K^2 + \frac{M_\eta^2}{3} \biggr) \nonumber\\
&+& \frac{M_\pi^3}{24 \pi m F^2}
\biggl( 3 + 3 g^2 - 4 C_3^q m + 8 C_4^q m \biggr)
+ \frac{M_K^3}{48 \pi m F^2}
\biggl( 3 + \frac{9}{2} g^2 + 8 C_4^q m \biggr)
+ \frac{g^2 M_\eta^3}{48 \pi m F^2} \biggr\} \nonumber\\
&+& 2 M_\pi^2 R \biggl( \bar D_{16}^q + \frac{1}{3} \bar D_{17}^q -
(1 + 2 g^2) L_{\pi\eta} \biggr) + 2 M_K^2 R \biggl( 2 \bar D_{16}^q
- \frac{1}{3} \bar D_{17}^q - \frac{1}{2} (1 + 3 g^2)
L_{K\eta}\biggr) \,.
\en
Substituting $g = \, \gaz/R$ and $m = \, \mnz/R^2$ we finally get:
\eq\label{gA_SU3} 
g_A &=& \gaz \biggl\{ 1
- \frac{\gaz^{\!\!\! 2}}{16 \pi^2 F^2 R^2} \biggl(M_\pi^2 + M_K^2
+ \frac{M_\eta^2}{3} \biggr) + \frac{M_\pi^3}{8 \pi \!\mnz \!\! F^2}
\biggl( R^2 + \gaz^{\!\!\! 2}
- \frac{4}{3} C_3^q \mnz + \frac{8}{3} C_4^q \mnz \biggr) \nonumber\\
&+& \frac{M_K^3}{16 \pi \!\mnz \!\! F^2} 
\biggl( R^2 + \frac{3}{2} \gaz^{\!\!\! 2}
+ \frac{8}{3} C_4^q \mnz \biggr)
+ \frac{\gaz^{\!\!\! 2} M_\eta^3}{48 \pi \!\mnz \!\! F^2} \biggr\} 
\nonumber\\
&+& 2 M_\pi^2 \, R \biggl( \bar D_{16}^q + \frac{1}{3} \bar D_{17}^q
- \frac{R^2 + 2 \gaz^{\!\!\! 2}}{R^3} \,
{\stackrel{\circ}L_{\pi\eta}} \biggr)
+ 2 M_K^2 \, R \biggl( 2 \, \bar D_{16}^q - \frac{1}{3} \bar D_{17}^q
- \frac{R^2 + 3 \gaz^{\!\!\! 2}}{2 R^3} \,
{\stackrel{\circ}L_{K\eta}}\biggr) \,,
\en
where
\eq
{\stackrel{\circ}L_{ab}}
= \frac{\gaz}{32 \pi^2 F^2} \ln\frac{M_a^2}{M_b^2} \,. 
\en

\subsection{Baryon octet semileptonic couplings}

Now we turn to the discussion of the vector and axial couplings
$g_V^{B_iB_j}$ and $g_A^{B_iB_j}$ (\ref{gV_gA}) governing the
semileptonic decays of the octet baryons.  Our results are
summarized in Table 3, and have a relatively simple structure.

In the case of the vector coupling $g_V^{B_iB_j}$, our results are
unchanged from those of the SU(3) limit in the case of the two
$\Delta S=0$ transitions, while in the case of the five $\Delta S=1$
transitions, our predictions are found by multiplying the simple
SU(3) limit forms by the common factor $1+\delta_V$.  Here
\eq
\delta_V = \delta f_1^{su} + \delta I_V^s + \delta f_1^{su} \delta
I_V^s\label{eq:bh}
\en
where the factors $\delta f_1^{su}$ and $\delta I_V^s$ have been
defined in Eqs.~(\ref{f1_delta}), and (\ref{delta_IV_IA}) and are
both second order in SU(3) breaking, in accord with the
Ademollo--Gatto theorem~\cite{Ademollo:1964sr}.

In the case of the axial coupling $g_A^{B_iB_j}$ the SU(3) symmetry
breaking is first order and, as derived in
Ref.~\cite{Ademollo:1964sr} and discussed e.g. in
Refs.~\cite{Garcia:1974cs,Song:1996mi}, can be described in terms of
an effective Lagrangian containing two SU(3) symmetric terms
proportional to the conventional couplings $D$ plus $F$ and four
first order SU(3) breaking terms proportional to the couplings 
$H_i$ $(i=1 \cdots 4)$: 
\eq\label{L_AG} 
{\cal L} &=& D \, \la \bar
B \gamma^\mu \gamma^5 \{a_\mu B \} \ra \, + \, F \, \la \bar
B \gamma^\mu \gamma^5 [a_\mu B ] \ra + \frac{H_1}{\sqrt{3}} \, \la \bar
B \gamma^\mu \gamma^5 B \{a_\mu \lambda_8 \} \ra +
\frac{H_2}{\sqrt{3}} \, \la \bar B \gamma^\mu \gamma^5
\{a_\mu \lambda_8 \} B \ra \nonumber\\[1mm]
&+& \frac{H_3}{\sqrt{3}} \, 
\la \bar B \gamma^\mu \gamma^5 a_\mu B \lambda_8
  - \bar B \gamma^\mu \gamma^5 \lambda_8 B a_\mu \ra
+ \frac{H_4}{\sqrt{3}} \,
 \biggl(  \, \la \bar B a_\mu \ra \gamma^\mu \gamma^5 \la B \lambda_8 \ra
+ \la \bar B \lambda_8 \ra \gamma^\mu \gamma^5 \la B a_\mu \ra \, \biggr)
\nonumber
\en
where
\eq
B =
\left(
\begin{array}{ccc}
\Sigma^0/\sqrt{2} + \Lambda/\sqrt{6}\,\, & \,\, \Sigma^+ \,\, & \, p \\
\Sigma^- \,\,  & \,\, -\Sigma^0/\sqrt{2}+\Lambda/\sqrt{6}\,\, & \, n \\
\Xi^-\,\,      & \,\, \Xi^0 \,\, & \, -2\Lambda/\sqrt{6}\\
\end{array}
\right) 
\en 
is the octet of baryon fields, while $a_\mu$ denotes
the external axial field \eq a_\mu = \left(
\begin{array}{ccc}
0 \,\, & \,\, a_\mu^{du} \,\, V_{ud} \,\,  & \,\, a_\mu^{su} \, V_{us} \\
0 \,\, & \,\, 0 \,\, & \,\, 0 \\
0 \,\, & \,\, 0 \,\, & \,\, 0 \\
\end{array}
\right) \,, 
\en 
with $V_{ud}$ and $V_{us}$ being the usual CKM matrix elements.

The axial semileptonic couplings of baryons are then expressed in
terms of the constants $D$, $F$ and $H_i$ as (see also
Refs.~\cite{Garcia:1974cs,Song:1996mi}):
\eq\label{gA_theory}
& &g_A^{np} = D + F + \frac{2}{3} (H_2 - H_3) \,, \nonumber\\[1mm]
& &g_A^{\Lambda p} = - \sqrt{\frac{3}{2}}
\biggl( F + \frac{D}{3} + \frac{1}{9} (H_1 - 2 H_2 - 3 H_3 - 6 H_4)
\biggr) \,, \nonumber\\[1mm]
& &g_A^{\Sigma^- n} = D - F - \frac{1}{3} (H_1 + H_3)
\,, \nonumber\\[1mm]
& &g_A^{\Sigma^- \Lambda} = \sqrt{\frac{2}{3}}
\biggl( D + \frac{1}{3} (H_1 + H_2 + 3 H_4)
\biggr) \,, \\[1mm]
& &g_A^{\Xi^- \Lambda} = \sqrt{\frac{3}{2}}
\biggl( F - \frac{D}{3} + \frac{1}{9} (2 H_1 - H_2 - 3 H_3 + 6 H_4)
\biggr) \,, \nonumber\\[1mm]
& &g_A^{\Xi^- \Sigma^0} = \sqrt{\frac{1}{2}}
\biggl( D + F - \frac{1}{3} (H_2 - H_3)
\biggr) \,, \nonumber\\[1mm]
& &g_A^{\Xi^0 \Sigma^+} = D + F - \frac{1}{3} (H_2 - H_3) \,.
\nonumber 
\en 
In our approach the axial couplings are expressed in
terms of the SU(3) symmetric contribution $g_A^{\rm SU_3} =
\frac{5}{3} g_1^{\rm SU_3} I_A$ and two symmetry breaking factors
$\delta_{A_1}$ and $\delta_{A_2}$ using the definitions (see also
Table 3): 
\eq
\frac{5}{3} g_1 I_A       &=& g_A^{\rm SU_3} ( 1 + \delta_{A_1} ) \,,\\
\frac{5}{3} g_1^{su}I_A^s &=& g_A^{\rm SU_3} ( 1 + \delta_{A_2} )
\,. 
\en 
The basic symmetry breaking pattern is then similar to that
in the case of the vector current --- the two $\Delta S=0$ transitions
are altered from their SU(3) values by one factor $1 + \delta_{A_1}$
while the five $\Delta S=1$ transitions are modified by a different
factor $1 + \delta_{A_2}$, where 
\eq\label{deltaA12} 
\delta_{A_1}
&=& \displaystyle\frac{\delta g_1}{g_1^{\rm SU_3}} \,,
\nonumber\\
\delta_{A_2} &=& \displaystyle\frac{\delta g_1^{su}}{g_1^{\rm SU_3}}
+ \displaystyle\frac{\delta I_A^s}{I_A} + \displaystyle\frac{\delta
g_1^{su}}{g_1^{\rm SU_3}} \displaystyle\frac{\delta I_A^s}{I_A} \,.
\en 
Note that both factors $\delta_{A_i}$ include not only 
leading --- ${\cal O}(m_s - \hat m)$ --- but also higher-order SU(3)  
breaking corrections.  In order to identify the effective couplings  
$D, F$ and $H_i$ in Lagrangian (\ref{L_AG}) we reduce the
expressions (\ref{deltaA12}) to the pieces first order in SU(3)
breaking.  Then the leading-order factors are given by [see
Eqs.~(\ref{delta_g1du}), (\ref{delta_g1su}) and (\ref{delta_IV_IA})]
\eq 
\delta_{A_1}^{(1)} &=& \displaystyle\frac{h_1}{g_1^{\rm SU_3}} 
(M_K^2 - M_\pi^2) \,,\\
\delta_{A_2}^{(1)} &=& \displaystyle\frac{h_2}{g_1^{\rm SU_3}}
(M_K^2 - M_\pi^2) - \displaystyle\frac{1 - I_A}{I_A} \delta \,. 
\en 
where the superscript $(1)$ indicates that we have truncated the
full expressions to include only the pieces first order in SU(3)
breaking.  Matching our results then for the axial couplings
$g_A^{B_i B_j}$ to the model--independent
predictions~(\ref{gA_theory}) we find 
\eq 
D = \frac{3}{2} F =
\frac{3}{5} g_A^{\rm SU_3} 
\en 
for the SU(3) symmetric contribution,
and 
\eq 
H_1 &=& \frac{1}{5} H_2 \, = \, \displaystyle\frac{1}{2} \,
I_A \, h_1 \, (M_K^2 - M_\pi^2) = - I_A \, h_2 \, (M_K^2 - M_\pi^2) 
+ g_1^{\rm SU_3} \,
(1 - I_A) \, \delta \label{H12} \,, \\[2mm]
H_3 &=& H_4 \, = \, 0 \label{H34} 
\en  
for the SU(3) breaking terms.
Therefore, to first order in SU(3) breaking the factors
$\delta_{A_1}^{(1)}$ and $\delta_{A_2}^{(1)}$ are not independent
and are related via $\delta_{A_1}^{(1)} = - 2 \delta_{A_2}^{(1)}$
which in terms of parameters of chiral Langrangian~(\ref{L_qU}) or
model-independent Lagrangian~(\ref{L_AG}) can be written as:
\eq\label{delta_A} 
\delta_A^{(1)} = \frac{2 H_1}{D} =
\displaystyle\frac{h_1}{g_1^{\rm SU_3}} (M_K^2 - M_\pi^2) = - \, 2
\, \biggl( \displaystyle\frac{h_2}{g_1^{\rm SU_3}} (M_K^2 - M_\pi^2)
- \displaystyle\frac{1 - I_A}{I_A} \, \delta \biggr) \, . 
\en 
Then using the relation (\ref{H12}) or (\ref{delta_A}) we deduce 
the following constraint involving the parameters of the chiral 
Lagrangian and the quantities defining the matrix elements of 
valence quarks (bare quark matrix elements):  
\eq 
G \, \frac{M_K^2 -
M_\pi^2}{(4 \pi F)^2} \ = \ \frac{1 - I_A}{I_A} \, \delta \,, \en
where \eq G = \frac{g}{g_1^{\rm SU_3}} \biggl( \frac{3}{2} +
\frac{23}{6} g^2 - \frac{10 \pi}{3} \bar M C_4^q \biggr) \,. 
\en 
The latter equation can be used to express the unknown quantity $\delta$ 
in terms of the parameters of the chiral Lagrangian~(\ref{L_qU}) ---
\eq\label{delta_constraint} 
\delta = \frac{G \, I_A}{1 - I_A}
\frac{M_K^2 - M_\pi^2}{(4 \pi F)^2} \,. 
\en 
Substituting Eqs.~(\ref{delta_IV_IA}) and (\ref{delta_constraint}) into
Eqs.~(\ref{eq:bh}) we can then in turn express $\delta_V^{(2)}$ (the
leading contribution to $\delta_V$ including second order SU(3)
breaking) in terms of the parameters of the chiral
Lagrangian~(\ref{L_qU}): 
\eq 
\delta_V^{(2)} &=& \delta f_1^{su} -
\frac{3}{2} \frac{G^2 \, I_A^2}{(1 - I_A) (1 + 3 I_A)} \frac{(M_K^2
- M_\pi^2)^2}{(4 \pi F)^4} \,. 
\en

\section{Numerical analysis}

Now we perform the numerical analysis of the vector and axial
couplings of quarks and baryons. First, 
we deduce constraints on the quark LEC's from the data on 
semileptonic decays of the baryon octet. Then we compare 
our results for the axial baryon couplings 
to the ones of baryon ChPT 
in the large--$N_c$ limit~\cite{FloresMendieta:2006ei}, 
obtained from a fit to the measured decays and 
ratios $g_A^{B_iB_j}/g_V^{B_iB_j}$. 

For the quark parameters in the
chiral limit we use $g = 0.9$ and $m = 420$~MeV, values fixed
previously in~\cite{Faessler:2005gd,Faessler:2006ky}.  Note that
these parameters are related to the corresponding nucleon quantities
$\gaz$ and $\mnz$ via the matching condition (\ref{m_mN}). In
particular, using the values $\gaz$ = 1.2, 1.2695 (data), 1.3 we
find for the nucleon mass in the chiral limit the results 
$\mnz$ = 746.7, 835.7, 876.3 MeV, respectively, which are consistent 
with the values deduced in the context of the baryon ChPT 
(see discussion in~\cite{Hemmert:2003cb,Procura:2006gq,Faessler:2005gd,%
Faessler:2006ky,Borasoy:1996bx,Frink:2005ru,Faessler:2007ee}).

First, we analyze the axial charges of the quark and the nucleon 
in SU(2) and SU(3), respectively.  In SU(2) the corresponding quantities in
terms of the LEC's $c_3^q$, $c_4^q$ and $d_{16}^q$ are given by
\eq
g_1    &=& 0.939 + 0.078 \, \alpha \,, \nonumber\\
\alpha &=& 0.195 \, (2 c_4^q - c_3^q) \, {\rm GeV}
+ \bar d_{16}^q \, {\rm GeV}^2
\en
and
\eq\label{gA_SU2_num}
g_A = \, \left\{
\begin{array}{l l}
1.251 + 0.104 \, \alpha
& \hspace*{.5cm} {\rm for}  \, \gaz = 1.2 \\
1.324 + 0.110 \, \alpha
& \hspace*{.5cm} {\rm for}  \,\gaz = 1.2695 \\
1.356 + 0.113 \, \alpha
& \hspace*{.5cm} {\rm for}  \, \gaz = 1.3
\end{array}
\right .
\en
Matching the expression for $g_A$ (\ref{gA_SU2_num}) to its  
experimental value we derive the following constraints on 
the SU(2) quark LEC's: 
\eq
\alpha = \, \left\{
\begin{array}{l l}
\hspace*{.3cm}
  0.178 & \hspace*{.5cm} {\rm for} \gaz = 1.2 \\
- 0.495 & \hspace*{.5cm} {\rm for} \gaz = 1.2695 \\
- 0.765 & \hspace*{.5cm} {\rm for} \gaz = 1.3
\end{array}
\right .
\en
Using the matching condition (\ref{matching_c3c4}), 
relating the combination $2 c_4^q - c_3^q$ of quark LEC's to the
corresponding ChPT LEC's~$c_3$ and $c_4$, and using the averaged
values of $c_3 = - 4.7$ GeV$^{-1}$ and $c_4 = 3.5$~GeV$^{-1}$
from~\cite{Fettes:1998ud} we estimate the LEC~$\bar d_{16}^q =
-1.957, -2.605, -2.868$~GeV$^{-2}$ (the corresponding ChPT LEC~$\bar
d_{16}$ is equal to $-2.469, -3.486, -3.931$~GeV$^{-2}$).  Note that
for the axial charge of the quark at one loop we find the values
$g_1 = 0.952, 0.9, 0.879$, respectively, which correspond to the 
values of $\gaz$ = 1.2, 1.2695, 1.3.

In SU(3) the corresponding results for the axial charges are
\eq
g_1   &=& 2.163 + 1.014 \, \beta\,, \nonumber\\
\beta &=& ( - 0.012 \, C_3^q +  0.563 \, C_4^q ) \, {\rm GeV}
 + ( \bar D_{16}^q - 0.147 \, \bar D_{17}^q ) \, {\rm GeV}^2 
\en
and
\eq\label{gA_SU3_num}
g_A = \,
\left\{
\begin{array}{l l}
2.884 + 1.352 \, \beta & \hspace*{.5cm} {\rm for} \gaz = 1.2 \\
3.051 + 1.430 \, \beta & \hspace*{.5cm} {\rm for} \gaz = 1.2695 \\
3.124 + 1.464 \, \beta & \hspace*{.5cm} {\rm for} \gaz = 1.3
\end{array}
\right .
\en
Matching the expression for $g_A$ (\ref{gA_SU3_num}) to its
experimental value we derive the following constraints on the SU(3)
quark LEC's:
\eq\label{beta}
\beta = \, \left\{
\begin{array}{l l}
- 1.194 & \hspace*{.5cm} {\rm for} \gaz = 1.2 \\
- 1.246 & \hspace*{.5cm} {\rm for} \gaz = 1.2695 \\
- 1.267 & \hspace*{.5cm} {\rm for} \gaz = 1.3
\end{array}
\right .
\en
Next we estimate the quark vector coupling $f_1^{su}$.
We determine
\eq
f_1^{su} = 1 + \delta f_1^{su} = 1.070 \,,
\en
{\it i.e.}, an SU(3)--breaking correction $\delta f_1^{su} = 7 \%$.

Next we extract information about the SU(3)--breaking parameters
$\delta_V^{(2)}$ and $\delta_A^{(1)}$
and find an additional constraint for the
linear combination of LEC's $C_3^q$, $C_4^q$ and $\bar D_{17}^q$
using data for the ratios $r^{B_iB_j} = g_A^{B_iB_j}/g_V^{B_iB_j}$.
Direct calculation of $\delta_V^{(2)}$ and $\delta_A^{(1)}$ gives:
\eq\label{Delta_V}
\delta_V^{(2)} = \left\{
\begin{array}{l l}
0.070 - 0.074 \, r_V^2
& \hspace*{.5cm} {\rm for} \gaz = 1.2 \\
0.070 - 0.103 \, r_V^2
& \hspace*{.5cm} {\rm for}  \gaz = 1.2695 \\
0.070 - 0.123 \, r_V^2
& \hspace*{.5cm} {\rm for} \gaz = 1.3 \\
\end{array}
\right  .
\en
and
\eq\label{Delta_A}
\delta_A^{(1)} =  - 0.136 \, r_A
\en
independent of the value for $\gaz$, where $r_V$ and $r_A$ are given by
\eq
r_V &=& \frac{1 - 0.935 \, C_4^q \, {\rm GeV}}{1 + 0.415 \, \gamma_1}\,, 
\nonumber\\[2mm]
r_A &=& \frac{1 + 0.449 \, \gamma_2}{1 + 0.415 \, \gamma_1}\,.
\en
Here $\gamma_1$ and $\gamma_2$ are the combinations of the quark LECs:
\eq\label{gamma12}
\gamma_1 &=& \bar D_{16}^q \, {\rm GeV}^2
- 0.311 \, ( C_3^q - 3 C_4^q ) \, {\rm GeV} \,, \nonumber\\
\gamma_2 &=& \bar D_{17}^q \, {\rm GeV}^2
- 1.400 \, ( 2 C_3^q - 3 C_4^q ) \, {\rm GeV}  \,.
\en
Matching our results for $r^{B_iB_j}$ to data~\cite{Yao:2006px} 
for the five semileptonic modes we have the following conditions 
involving the parameters $\delta_V$ and $\delta_A$:
\eq\label{gagv_ratios}
& &r^{np} = g_A = g_A^{\rm SU_3} \, (1 + \delta_A^{(1)})
= 1.2695 \pm 0.0029 \,, \nonumber\\[1mm]
& &r^{\Lambda p} = \frac{3 g_A^{\rm SU_3}}{5}
\,  \frac{1 - \frac{1}{2}\delta_A^{(1)}}{1 + \delta_V^{(2)}}
= 0.718 \pm 0.015 \,,
\nonumber\\[1mm]
& &r^{\Sigma n} = - \frac{g_A^{\rm SU_3}}{5} \,
\frac{1 - \frac{1}{2}\delta_A^{(1)}}{1 + \delta_V^{(2)}}
= -0.34 \pm 0.017\,, \\
& &r^{\Xi^- \Lambda} = \frac{g_A^{\rm SU_3}}{5} \,
\frac{1 - \frac{1}{2}\delta_A^{(1)}}{1 + \delta_V^{(2)}}
 = 0.25 \pm 0.05\,, \nonumber\\[1mm]
& &r^{\Xi^0 \Sigma^+} = g_A^{\rm SU_3} \,
\frac{1 - \frac{1}{2}\delta_A^{(1)}}
{1 + \delta_V^{(2)}} = 1.20 \pm 0.04 \pm 0.03\,. \nonumber
\en
Restricting to the central values of the data, we deduce the  
following constraints on $\delta_V^{(2)}$ and $\delta_A^{(1)}$:
\eq
g_A^{\rm SU_3} \, (1 + \delta_A^{(1)})
&=& 1.2695 \hspace*{1cm}
 {\rm from \ the} \ \ n \to p \ \ {\rm transition} \label{dAV1} \\[3mm]
g_A^{\rm SU_3} \,
\frac{1 - \frac{1}{2}\delta_A^{(1)}}{1 + \delta_V^{(2)}}
&=& \, \left\{
\begin{array}{l l l}
1.197 & \hspace*{.5cm}
{\rm from \ the} \ \  \Lambda \to p & {\rm transition}  \\
1.700 & \hspace*{.5cm}
{\rm from \ the} \ \ \Sigma^- \to n & {\rm transition}  \\
1.250 & \hspace*{.5cm}
{\rm from \ the} \ \ \Xi^- \to \Lambda & {\rm transition}  \\
1.200 & \hspace*{.5cm}
{\rm from \ the} \ \ \Xi^0 \to \Sigma^+ & {\rm transition}
\end{array}
\right . \label{dAV2}
\en
The three modes ($\Lambda \to p$,
$\Xi^- \to \Lambda$, $\Xi^0 \to \Sigma^+$) are quite consistent with
each other.  Future, more precise data for the $\Sigma^- \to n$ mode
will probably yield a smaller value for $r^{\Sigma n}$. As already stated
above, in order to get a better quantitative agreement with experiment 
we plan to go beyond the simple SU(6) quark model. Then we intend to
evaluate the valence quark matrix elements (see discussion in Sec.II.D)
in a fully relativistic quark model based on a specific scenario about
hadronization and confinement of quarks inside the
baryon~\cite{Faessler:2007qm}.  Roughly speaking, instead of the
trivial identities (\ref{dAV1}) and (\ref{dAV2})
involving only two SU(3) breaking
parameters $\delta_V$ and $\delta_A$ we will derive more general
identities involving additional symmetry breaking parameters.

Using Eq.~(\ref{dAV1}) we derive the following constraint 
on the quark LECs:
\eq\label{gamma12_con}
\gamma_1 - 0.147 \gamma_2 =
\left\{
\begin{array}{l l}
- 1.144
& \hspace*{.5cm} {\rm for} \gaz = 1.2 \\
- 1.195
& \hspace*{.5cm} {\rm for} \gaz = 1.2695 \\
- 1.216
& \hspace*{.5cm} {\rm for} \gaz = 1.3 \\
\end{array}
\right .
\en
Next, using the typical value $\simeq 1.2$ for the ratios
in Eq.~(\ref{dAV2}) we deduce the following
constraints on the linear combinations of quark LEC's:
\eq\label{sigma}
\left .
\begin{array}{l l}
\gamma_1 + 0.078 \, \gamma_2 - 0.130 \, C_4^q \, {\rm GeV} = - 1.787
& \hspace*{.5cm} {\rm for} \gaz = 1.2\,, \\
\gamma_1 + 0.079 \, \gamma_2 - 0.174 \, C_4^q \, {\rm GeV} = - 1.899
& \hspace*{.5cm} {\rm for} \gaz = 1.2695\,, \\
\gamma_1 + 0.080 \, \gamma_2 - 0.205 \, C_4^q \, {\rm GeV} = - 1.962
& \hspace*{.5cm} {\rm for} \gaz = 1.3\,.  \\
\end{array}
\right .
\en
Finally, using two equations
(\ref{gamma12_con}) and (\ref{sigma}) on four LECs
$C_3^q$, $C_4^q$, $\bar D_{16}^q$ and $\bar D_{17}^q$
we can express two of them (e.g. $\bar D_{16}^q$ and $\bar D_{17}^q$)
through the other two ($C_3^q$ and $C_4^q$) as:

For $\gaz = 1.2$
\eq\label{constraint_g12}
\left .
\begin{array}{l}
\bar D_{16}^q  = - 1.565 \, {\rm GeV}^{-2}
        + 0.311 \, ( C_3^q - 2.727 \, C_4^q )  \, {\rm GeV}^{-1} \,, \\
\bar D_{17}^q  = - 2.862 \, {\rm GeV}^{-2}
        + 2.800 \, ( C_3^q - 1.294 \, C_4^q )  \, {\rm GeV}^{-1} \,. \\
\end{array}
\right .
\en

For $\gaz = 1.2695$
\eq\label{constraint_g12695}
\left .
\begin{array}{l}
\bar D_{16}^q  = - 1.652 \, {\rm GeV}^{-2}
        + 0.311 \, ( C_3^q - 2.637 \, C_4^q )  \, {\rm GeV}^{-1} \,, \\
\bar D_{17}^q  = - 3.111 \, {\rm GeV}^{-2}
        + 2.800 \, ( C_3^q - 1.225 \, C_4^q )  \, {\rm GeV}^{-1} \,. \\
\end{array}
\right .
\en

For $\gaz = 1.3$
\eq\label{constraint_g13}
\left .
\begin{array}{l}
\bar D_{16}^q  = - 1.699 \, {\rm GeV}^{-2}
        + 0.311 \, ( C_3^q - 2.573 \, C_4^q )  \, {\rm GeV}^{-1} \,, \\
\bar D_{17}^q  = - 3.283 \, {\rm GeV}^{-2}
        + 2.800 \, ( C_3^q - 1.177 \, C_4^q )  \, {\rm GeV}^{-1} \,. \\
\end{array}
\right .
\en
Note, that the constraint (\ref{beta}) on the SU(3) quark LECs
was obtained without dropping the higher--order terms in SU(3)
breaking, while the constraints 
(\ref{constraint_g12})--(\ref{constraint_g13}) were derived
using the approximation for SU(3) breaking terms
$\delta_V \to \delta_V^{(2)}$
and $\delta_{A_i} \to \delta_{A_i}^{(1)}$ restricting
to their leading terms.

Finally, for completeness we also present numerical results for the axial 
couplings at values of $\gaz = 1.2$ and $C_4^q = 1.07$ GeV$^{-1}$. The other 
three LEC's $C_3^q$, $\bar D_{16}^q$ and $\bar D_{17}^q$ are then 
constrained as: 
\seq
\eq 
\bar D_{16}^q  - 0.311 \, C_3^q  = - 1.668 \, {\rm GeV}^{-2} \, \\
\bar D_{17}^q  - 2.800 \, C_3^q  = - 6.739 \, {\rm GeV}^{-2} \,. 
\en 
\sen 
Predictions for $g_A^{B_iB_j}$ of different semileptonic modes 
are given in Table 4. We also indicate the results
of heavy baryon ChPT in the large--$N_c$ limit~\cite{FloresMendieta:2006ei}. 
In Table 5 we additionally present our results for the semileptonic 
decay widths of hyperons, which are calculated using the 
expression~\cite{Pietschmann:1974ap}
at order ${\cal O}((m_{B_i}-m_{B_j})^6)$ 
and without inclusion of radiative corrections:   
\eq\label{decay_width}  
\Gamma(B_i \to B_j + l + \nu_l) = \frac{G_F^2}{60 \pi^3} 
|V_{\rm CKM}|^2 (\Delta m)^5 (1 - 3 \delta) 
\biggl( (g_V^{B_iB_j})^2 + 3 (g_A^{B_iB_j})^2 \biggr) \ r(x)\,. 
\en 
In the last expression we have $\Delta m = m_{B_i} - m_{B_j}$, 
$\delta = (m_{B_i} - m_{B_j})/(m_{B_i} + m_{B_j})$, 
$G_F = 1.16637 \times 10^{-5}$ GeV$^{-2}$ is the Fermi coupling constant.
For the corresponding CKM matrix elements 
$V_{\rm CKM} = V_{ud}$ or $V_{us}$ 
we use the central values from~\cite{Yao:2006px}: 
$V_{ud} = 0.97377$ and $V_{us} = 0.225$. Here $r(x)$ is the function 
which takes into account the charged lepton mass $m_l$: 
\eq 
r(x) = \sqrt{1 - x^2} \biggl(1 - \frac{9}{2} x^2 - 4 x^4 \biggr) 
+ \frac{15}{4} x^4 \ln\frac{1 +  \sqrt{1 - x^2}}{1 - \sqrt{1 - x^2}} 
\en 
where $x = m_l/\Delta m$ and $r(0) = 1$. 

\section{Summary}

In this paper we have analyzed the semileptonic vector and axial 
quark coupling constants using a manifestly Lorentz covariant chiral  
quark approach up to order ${\cal O}(p^4)$ in the two-- and 
three--flavor picture. The resulting quark couplings were then  
used in the evaluation of the corresponding hadronic couplings which   
govern semileptonic transitions between baryon octet states.  In the  
calculation of baryon matrix elements we utilized a general ansatz  
for the spatial form of the quark wave function, without referring to   
any specific realization of baryon hadronization and confinement.  
Matching physical amplitudes, calculated within our approach, to the 
model--independent predictions of baryon chiral perturbation theory 
(ChPT) allowed us to deduce the relations between the chiral quark 
parameters and those of baryon ChPT. 

Our main results can be summarized as follows:

-- Evaluating the chiral and SU(3) symmetry--breaking corrections to
the semileptonic vector and axial quark coupling constants, we
determined that the SU(3) symmetry--breaking correction to the vector
coupling $f_1^{su}$, governing the $s \to u$ quark flavor transition, 
is positive and equal to 7\%;

-- Performing the matching to ChPT we reproduced the analytical
result for the nucleon axial charge $g_A$ in SU(2). We also 
determined the expression for $g_A$ in SU(3);

-- We derived results for the vector and axial couplings governing
the semileptonic decays of the baryon octet, revealing both chiral
and SU(3) symmetry--breaking corrections; 

-- We presented a numerical analysis of the calculated quantities and
derived constraints on the parameters of the chiral quark Lagrangian
(LEC's) using experimental data for $g_A$ and the ratios $r^{B_iB_j}
= g_A^{B_iB_j}/g_V^{B_iB_j}$. We also gave estimates for the semileptonic 
decay widths of hyperons. 

In future we plan to improve the quantitative determination of 
the valence quark effects by resorting to a relativistic quark 
model~\cite{Ivanov:1996pz,Faessler:2006ky}, describing the internal 
quark dynamics. This procedure will allow us to give predictions for 
all six form factors showing up in the matrix elements of the 
semileptonic decays of the baryon octet. With the explicit form factors 
and with additional radiative corrections included we intend to give 
accurate predictions for the corresponding decay widths and asymmetries.

\begin{acknowledgments}

This work was supported by the DFG under contracts FA67/31-1 and
GRK683. BRH is supported by the US National Science Foundation under
Grant No. PHY 05-53304. This research is also part of the EU
Integrated Infrastructure Initiative Hadronphysics project under
contract number RII3-CT-2004-506078 and President grant of Russia
"Scientific Schools"  No. 871.2008.2. 

\end{acknowledgments}

\appendix\section{Contributions of different diagrams to the
vector and axial quark couplings} 

In this Appendix we discuss the contributions of the various graphs
in Figs.1 and 2 to the vector and axial couplings with different
flavor structures. The separate contributions of these graphs to the
vector and axial couplings are listed in Tables 1 and 2,
respectively. We use the following notations. The quark charge 
matrix ${\cal Q} = {\rm diag}\{2/3, -1/3\}$ in SU(2) and $Q = {\rm
diag}\{2/3, -1/3, -1/3\}$ in SU(3); the unit $2 \times 2$ matrix
${\cal I} = {\rm diag}\{1, 1\}$ and $3 \times 3$ matrix $I = {\rm
diag}\{1, 1, 1\}$. All further flavor matrices are expressed in
terms of the charge, unit, Pauli ($\tau_i$) and Gell-Mann
($\lambda_i$) matrices: 
$$\tau_{ud} = \frac{1}{2} (\tau_1 + i \tau_2)\,, \hspace*{1cm}
\lambda_{ud} = \frac{1}{2} (\lambda_1 + i \lambda_2)\,, \hspace*{1cm}
\lambda_{us} = \frac{1}{2} (\lambda_4 + i \lambda_5)\,,$$ 
$$\beta_\pi^Q = \frac{3}{4}Q + \frac{1}{4}I + \frac{3}{4}\lambda_3\,,
\hspace*{1cm}
\beta_K^Q = \frac{5}{2}Q - \frac{1}{6}I - \frac{1}{2}\lambda_3\,,
\hspace*{1cm}
\beta_\eta^Q = \frac{3}{4}Q - \frac{1}{12}I - \frac{1}{4}\lambda_3\,,$$
$$\beta_\pi = \frac{9}{4}\,, \hspace*{1cm}
\beta_K = \frac{3}{2}\,, \hspace*{1cm} \beta_\eta = \frac{1}{4}\,,$$
$$\gamma_\pi = \frac{9}{8}\,, \hspace*{1cm}
\gamma_K = \frac{9}{4}\,, \hspace*{1cm} \gamma_\eta = \frac{5}{8}\,,$$
$$\lambda_\pi^b = \lambda_\pi^c = \lambda_\pi^d = \lambda_\pi^f =
\lambda_3\,, \hspace*{1cm}
\lambda_K^b = \lambda_K^c = \lambda_K^d = \lambda_K^f =
3Q - \lambda_3\,, \hspace*{1cm}
\lambda_\eta^b = \lambda_\eta^c = \lambda_\eta^d = \lambda_\eta^f = 0\,,$$
$$\lambda_\pi^e = Q + \frac{1}{3}I - \lambda_3\,, \hspace*{1cm}
  \lambda_K^e = -\frac{8}{3}Q - \frac{2}{9}I + \frac{4}{3} \lambda_3\,,
\hspace*{1cm}
\lambda_\eta^e = Q - \frac{1}{9}I - \frac{1}{3} \lambda_3\,.$$

We introduce the functions $R_P, R_P^i$, $T_P^i$, $I_{ab}$ and $J_{ab}$:  
\eq\label{fun_R_P}
R_P &=& \frac{g^2}{F^2} \Delta_P + \frac{g^2 M_P^2}{24 \pi^2 F^2}
\biggl\{ 1 - \frac{3 \pi}{2} \mu_P \biggr\}\,, \nonumber\\
R_P^b &=& \frac{3 g^2}{2 F^2}\Delta_P
+ \frac{g^2 M_P^2}{16 \pi^2 F^2} \biggl\{ 1 - \frac{5 \pi}{2} \mu_P
\biggr\}\,, \nonumber\\
R_P^e &=& \frac{3}{4} R_P\,, \nonumber\\
R_P^f &=& \frac{g^2 M_P^2}{16 \pi F^2} \mu_P \,, \nonumber\\
T_P^d &=& \frac{g^3}{4F^2} \Delta_P + \frac{g^3 M_P^2}{32 \pi^2 F^2}
\biggl\{ 1 - \frac{\pi}{2} \mu_P \biggr\} \,, \nonumber\\
T_P^e &=& \frac{g}{8 \pi F^2} M_P^2 \mu_P \,, \nonumber\\
T_P^f &=& - \frac{g}{6 \pi F^2} M_P^3  \,, 
\en 
where 
$\mu_P = \frac{M_P}{m}$, and 
\eq 
I_{ab} = \frac{3g^2}{4 F^2} 
\biggl[ \frac{3}{2} \frac{\Delta_{a} M_a^2 
- \Delta_{b} M_b^2}{M_a^2 - M_b^2} 
+ \frac{M_a^2 + M_b^2}{64 \pi^2} - \frac{M_a^3 + M_b^3}{8 \pi m} 
- \frac{M_a^2 M_b^2}{8 \pi m (M_a + M_b)} \biggr] 
\en 
and 
\eq
J_{ab} = \frac{3}{8 F^2} \biggl[ 
\frac{\Delta_{a} M_a^2 - \Delta_{b} M_b^2}{M_a^2 - M_b^2} 
- \frac{M_a^2 + M_b^2}{32 \pi^2} \biggr] 
\en 
Below we discuss the vector couplings in detail.

\vspace*{.3cm}

1. Electric charges: Summing up the individual contributions of the 
graphs in Fig.1 to the electric charges we find 
\eq
f_1^{\cal Q} =
{\cal Q} (1 - \frac{9}{4} R_\pi) + \tau_3 ( R_\pi^b + R_\pi^f ) +
\frac{1}{2} ({\cal I} - \tau_3) R_\pi^e \en in SU(2) and \eq f_1^Q =
Q + \sum\limits_P ( - \beta_P^Q R_P + \lambda_P^b R_P^b
                            + \lambda_P^e R_P^e + \lambda_P^f R_P^f )
\en
in SU(3).

Using the identities $R_P^b + R_P^f = \frac{3}{2}R_P$ 
and $\beta_P^Q = \frac{3}{2} \lambda_P^b  + \frac{3}{4} \lambda_P^e$ 
we verify electric charge conservation --- 
$f_1^{\cal Q} = {\cal Q}$ in SU(2) and $f_1^Q = Q$ in SU(3).

\vspace*{.3cm}

2. The isotopic charge $f_1/2$ and the $d \to u$ transition 
vector coupling $f_1^{du}$ are given by the expressions: 
\eq
f_1 = f_1^{du}
= 1 - \frac{9}{4} R_\pi + 2 ( R_\pi^b + R_\pi^f ) - R_\pi^e
\en
in SU(2) and
\eq
f_1 = f_1^{du} = 1 - \sum\limits_P \beta_P R_P + 2
(R_\pi^b + R_\pi^f) + R_K^b + R_K^f - R_\pi^e + \frac{1}{3}
R_\eta^e
\en
in SU(3).

Again, using identities involving the functions $R_P^i$ we arrive at
\eq
f_1 = f_1^{du} = 1
\en
in both the two-- and three--flavor pictures.

\vspace*{.3cm}

3. The $s \to u$ transition vector coupling $f_1^{su}$: 
The coupling $f_1^{su}$ is finite but contains pieces  
${\cal O}((M_K-M_\pi)^2)$ and ${\cal O}((M_K-M_\eta)^2)$ 
which are of second order in SU(3) symmetry breaking.  
The Ademollo--Gatto theorem (AGT) protects the
$f_1^{su}(0)$ from {\it first}--order symmetry breaking corrections.
Moreover, the AGT holds independently for two sets of diagrams --- for
set I, including the diagrams of Fig.1(a), (b), (e), and (f) and for
set II, including the diagrams of Fig.1(c) and (d). In the our
derivation we use the identity
\eq
\frac{\Delta_a M_a^2 - \Delta_b M_b^2}{M_a^2 - M_b^2} 
= 2 \lambda_a ( M_a^2 + M_b^2) + \frac{1}{16 \pi^2} 
\frac{M_b^4}{M_a^2 - M_b^2} {\rm ln}\frac{M_a^2}{M_b^2} \,. 
\en
Then the results for set I and set II are:
\eq
f_1^{su; I} = \sum\limits_{i = a, b, e, f} f_1^{su; (i)} =
1 - \frac{9 g^2}{16} ( H_{\pi K} + H_{\eta K} + G_{\pi K} + G_{\eta K} )
\en
and
\eq
f_1^{su; II} = \sum\limits_{i = c, d} f_1^{su; (i)}
= - \frac{3}{16} ( H_{\pi K} + H_{\eta K} ) \,, 
\en
where the functions 
$H_{ab} = O ( (M_a^2 - M_b^2)^2 )$ and 
$G_{ab} = O ( (M_a^2 - M_b^2)^2 )$ 
defined in Eq.~(\ref{fun_HG}) of Sec.II.C.

The final result for the $s \to u$ quark transition
vector coupling is:
\eq
f_1^{su} = f_1^{su; I} + f_1^{su; II} =
1 - \frac{3}{16} \biggl( (1 + 3 g^2) ( H_{\pi K} + H_{\pi K} )
+ 3 g^2 ( G_{\pi K} + G_{\pi K} ) \biggr) \,. 
\en

\section{Two-body operators} 

The diagrams contributing to the two--body vector and axial quark  
transition operators up to fourth order are displayed in Figs.3 and Fig.4. 
First, let us discuss the diagrams in Figs.3(a)-(e) and 4(a)-(e). 
Note, the diagrams in Figs.3(c,d) and 4(c,d) are generated by an insertion 
of the two--body mass counterterm due to one--meson exchange, 
which is given by the four--quark operator 
\eq 
O_{\rm ct}(x,y) = \frac{g^2 m^2}{2F^2} 
\sum\limits_{i=1}^8 \bar q(x) \gamma_5 \lambda_i  q(x) 
\Delta_{ij}(x-y) \bar q(y) \gamma_5 \lambda_j  q(y) 
\en 
where 
\eq 
\Delta_{ij}(x-y) = \la 0 | T \phi_i(x) \phi_j(y) | 0 \ra = 
\delta_{ij} \int \frac{d^4 k}{(2\pi)^4 i} 
\frac{e^{-ik(x-y)}}{M_i^2 - k^2} 
\en 
is the meson propagator. Writing down the expressions for the 
diagrams in Figs.3(a)-(e) in the momentum space it is easy to show 
that the contribution of the diagrams in Figs.3(a), 3(b) and 3(e) is 
exactly equal to the contribution of the diagrams in Figs.3(c) and 3(d) 
but with opposite sign. Therefore, their total contribution vanishes. 
Such cancellation guarantees the charge conservation and excludes a 
double--counting of the one--meson exchange corrections. 
The diagram in Fig.3(f) 
does not contribute to the time component of the vector current (only to the 
spatial component), therefore we have no contribution to the baryon 
vector couplings from the two--body operators displayed in Fig.3. 

In the case of the two--body axial diagrams, the diagrams in 
Figs.4(a)-(e) do not cancel each other. Their total contribution 
in momentum space is given by 
\eq\label{axial_contr}  
\frac{m g}{4 F^2} (g^2 - 1 )\sum\limits_{i=1}^8 
\frac{1}{M_i^2 - k^2} 
\bar u(p_1^\prime) \, [ \lambda_A\,, \lambda_i ] \gamma^\mu \, u(p_1) \ 
\bar u(p_2^\prime) \, \lambda_i \gamma^5 \, u(p_2) + (1 \leftrightarrow 2) 
\en  
where $u(p_i)$ and $\bar u(p_i^\prime)$ are the quark spinors, 
$\lambda_A$ is the flavor matrix corresponding to the axial 
quark flavor exchange. One can see, that the contribution (\ref{axial_contr}) 
vanishes for $g = 1$. 
The other two diagrams in Fig.4(f) and 4(g) are generated by 
the one--body and two--body Lagrangians and they contribute to the 
axial couplings of the baryon octet. Note, that nonvanishing  
two-body operators corresponding to the meson exchange can be 
simplified. One can do the expansion of the meson propagators 
in powers of meson masses $M$ as 
\eq 
\frac{1}{\Lambda^2 - M^2} = \frac{1}{\Lambda^2} 
\biggl( 1 + \frac{M^2}{\Lambda^2} + {\cal O}(M^4) \biggr) \,. 
\en 
$\Lambda$ is a free parameter representing an averaged exchanged 
momenta between quarks, and we remove the infrared-regular parts proportional 
to the $1/\Lambda^N$, i.e. they do not contain powers of meson masses. 
Numerical analysis of the contributions of the two--body diagrams will 
be done in future. Let us stress again, that the vector couplings of the 
baryons do not receive contributions from the two--body quark operators 
(see diagrams in Fig.3), while the axial couplings receive the corrections 
quadratic in meson masses. It will not damage the nonanalytical chiral 
corrections derive in the one--body approximation (see Sec.~III) and only 
will redefine the expressions for the quadratic corrections. Note, that 
such change of the quadratic chiral corrections will be consistent with ChPT 
due to the matching condition involving additional two--body quark LEC's.

\newpage

\begin{center}
\epsfig{figure=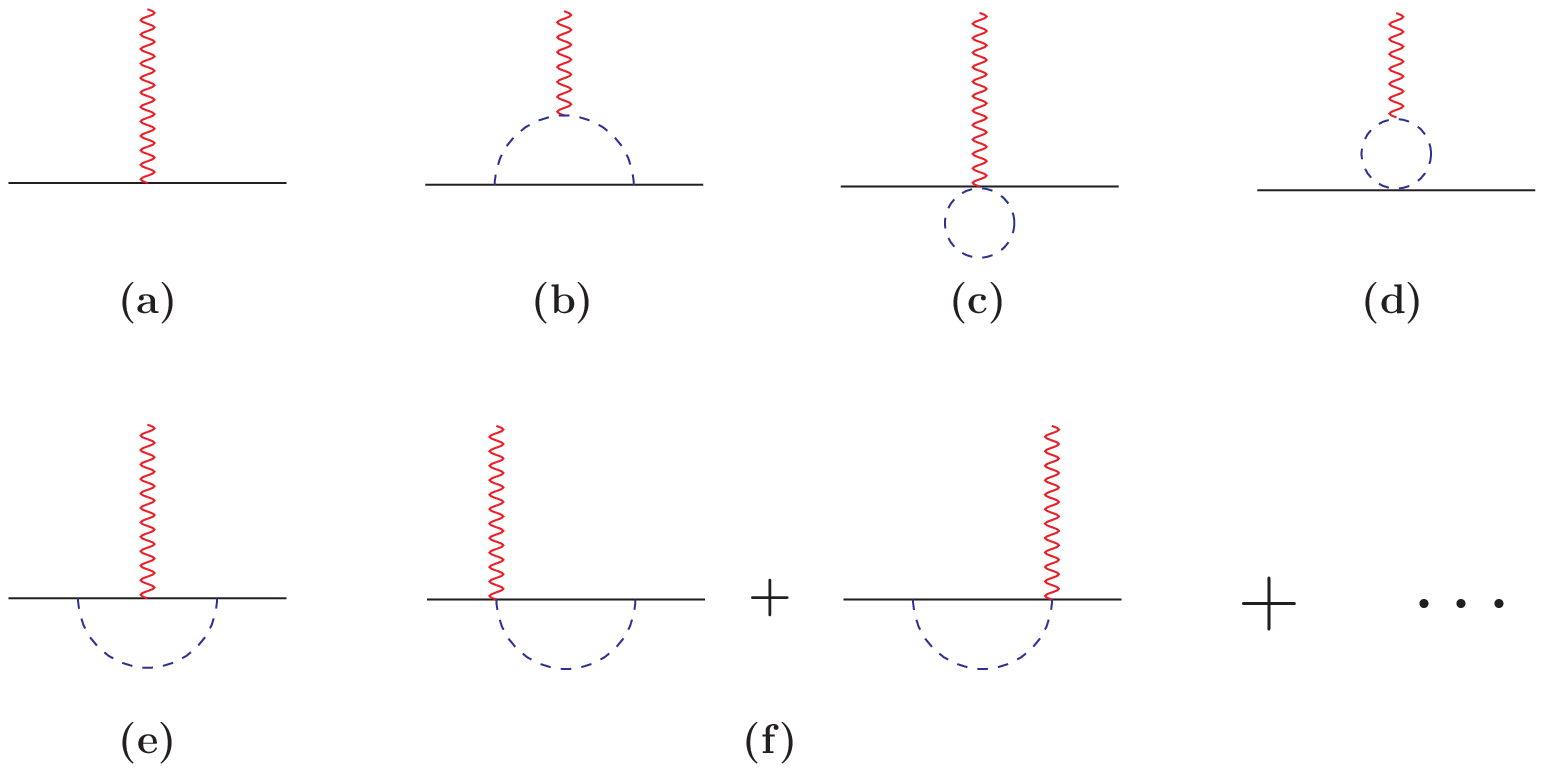,scale=.9}
\end{center}

\vspace*{0.5cm}

\noindent
{\bf Fig. 1.} {\em Diagrams contributing to the one--body vector quark  
transition operator up to fourth order. Solid, dashed and wiggly 
lines refer to quarks, pseudoscalar mesons 
and the external vector field, respectively.  

\label{fig1}}

\vspace*{1cm}

\begin{center}
\epsfig{file=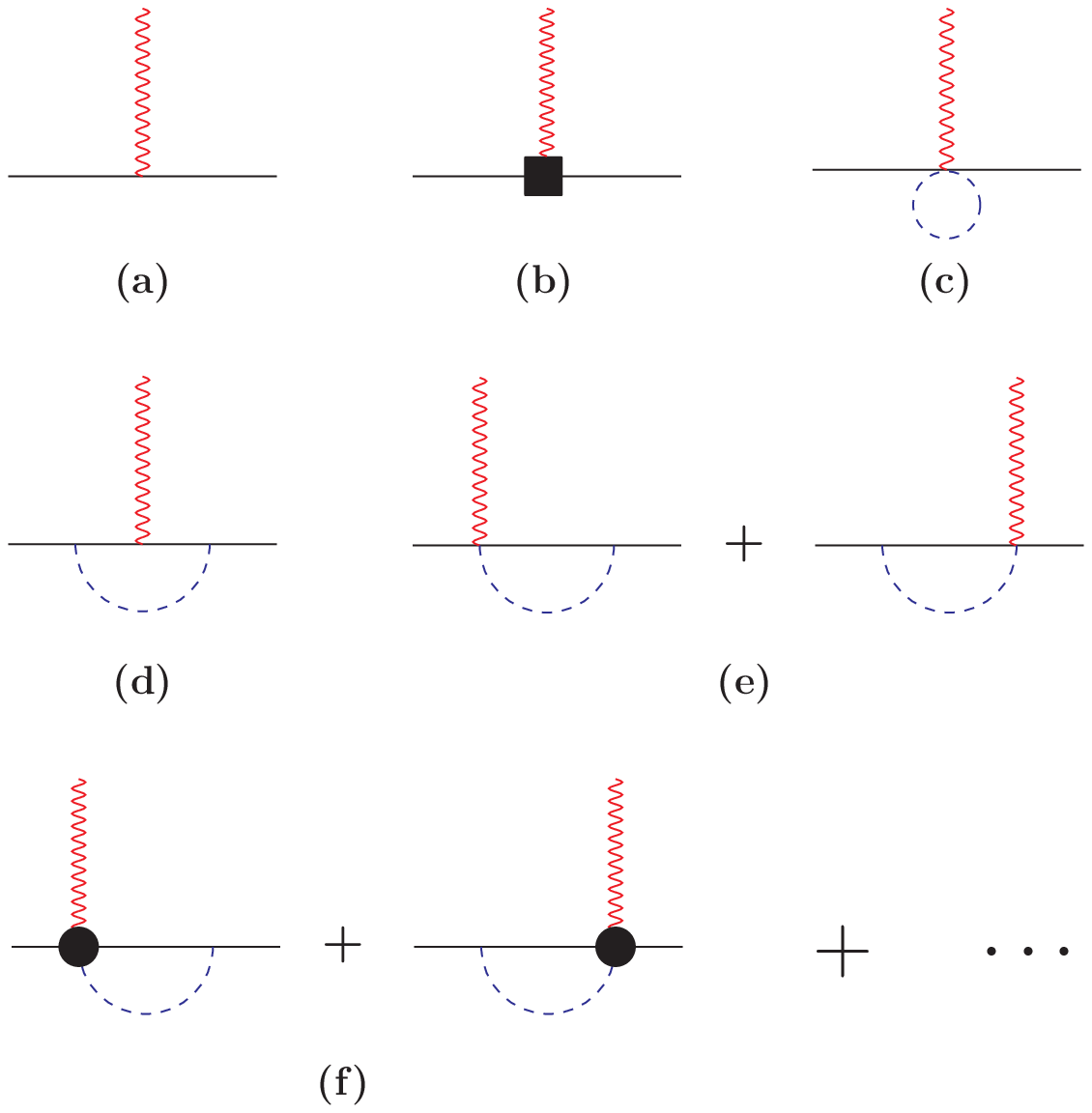,scale=.9}
\end{center}

\vspace*{0.5cm}

{\bf Fig. 2.} {\em Diagrams contributing
to the one--body axial quark transition operator up to fourth order.
Solid, dashed and wiggly lines refer to quarks, pseudoscalar mesons
and the external axial field, respectively. Vertices denoted by a black
filled circle and box correspond to insertions from the second
and third order chiral Lagrangian.
\label{fig2}}

\newpage

\begin{center}
\epsfig{figure=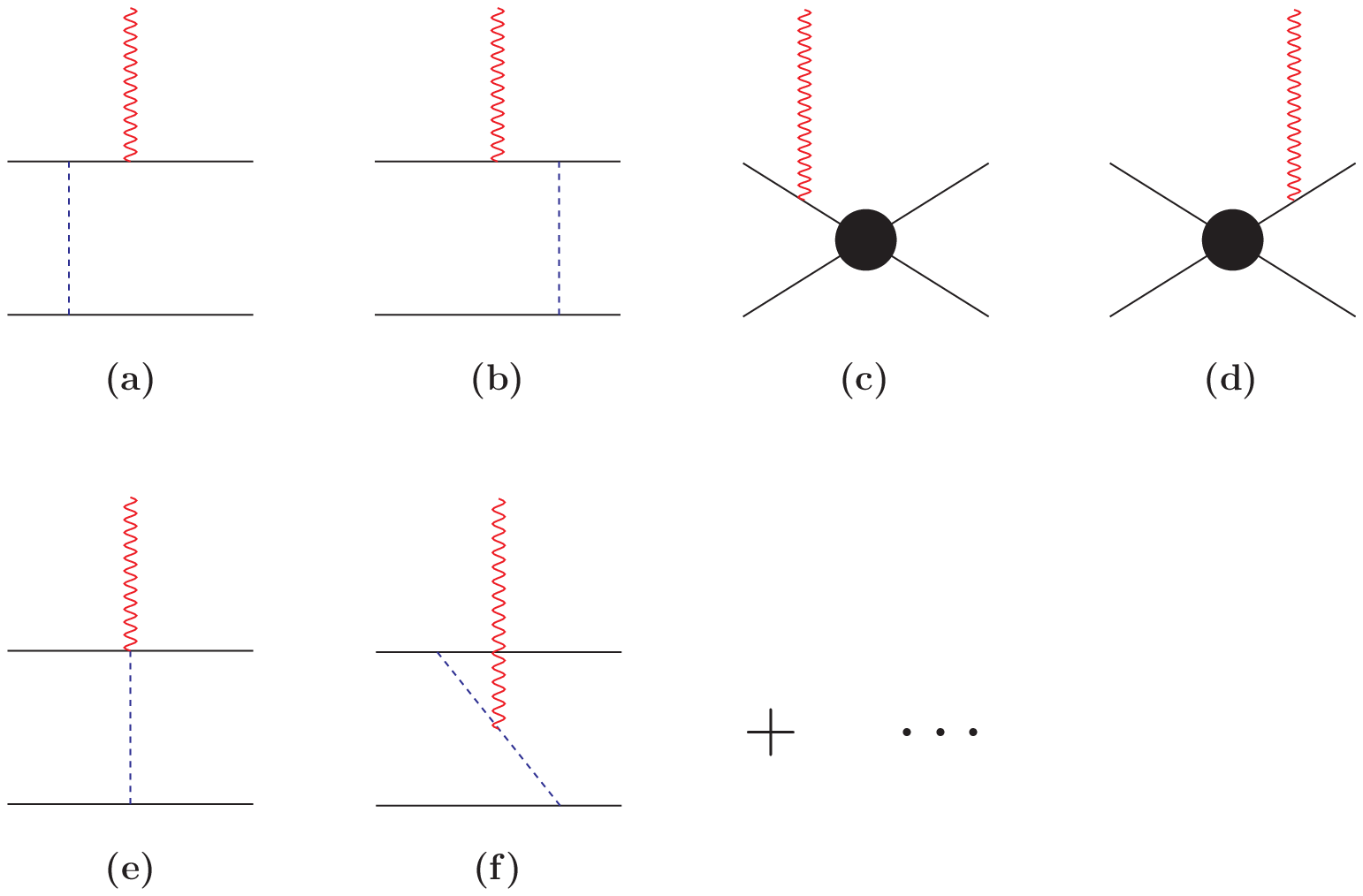,scale=.85}
\end{center}

\vspace*{0.1cm}

\noindent 
{\bf Fig. 3.} {\em Diagrams contributing to the two--body vector quark  
transition operator up to fourth order. Solid, dashed and wiggly 
lines refer to quarks, pseudoscalar mesons and the external vector field, 
respectively. The vertex denoted by a big black filled circle corresponds 
to insertion of the two-body mass counterterm due to one--meson exchange. 
\label{fig3}}

\vspace*{1cm}

\begin{center}
\epsfig{file=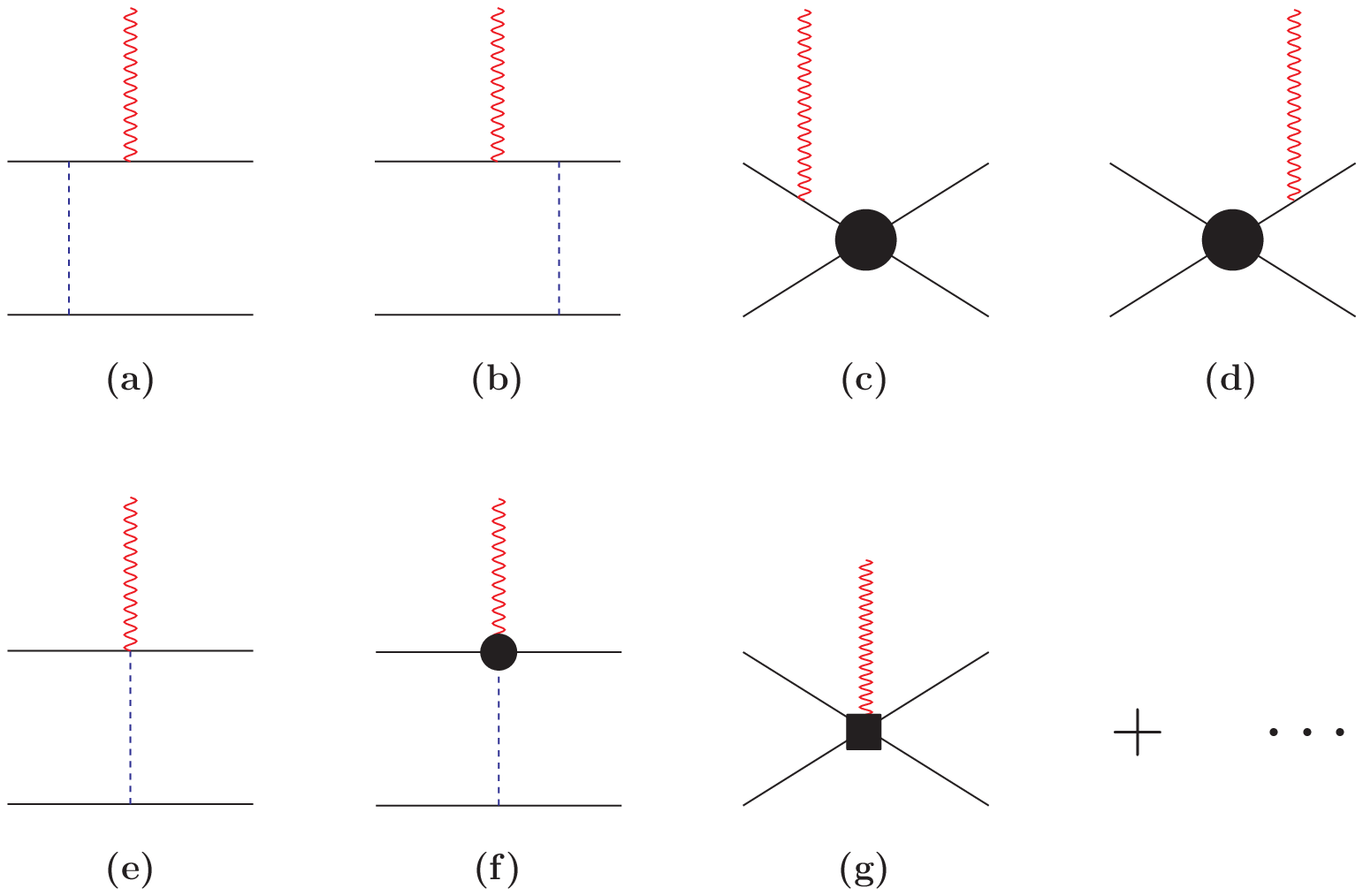,scale=.85}
\end{center}

\vspace*{0.1cm}

{\bf Fig. 4.} {\em Diagrams contributing
to the two--body axial quark transition operator up to fourth order.
Solid, dashed and wiggly lines refer to quarks, pseudoscalar mesons
and the external axial field, respectively. Vertices denoted by 
a big (small) black filled circle and box correspond to insertions 
of the two-body mass counterterm due to one--meson exchange, 
from the second and third order chiral Lagrangian. 
\label{fig4}}

\newpage

\noindent
{\bf Table 1.} Contribution of different diagrams in Fig.1 to
the electric charge $f_1^{\cal Q}$ [in SU(2)] and $f_1^Q$ [in SU(3)],
isotopic (vector) charge $f_1 \tau_3/2$ [in SU(2)]
and $f_1 \lambda_3/2$ [in SU(3)], vector coupling
($d \to u$ flavor transition) $f_1^{du} \tau_{ud}$ [in SU(2)] and
$f_1^{du} \lambda_{ud}$ [in SU(3)] and vector coupling ($s \to u$
flavor transition) $f_1^{su} \lambda_{us}$. The contribution of diagram
in Fig.1(a) is multiplied by the $Z$--factor.

\begin{center}
\def\arraystretch{2}
\hspace*{-.7cm}
\begin{tabular}{|c|c|c|c|c|c|c|}
\hline
 Coupling           &  Fig.1(a) &  Fig.1(b) &  Fig.1(c)
                    &  Fig.1(d) &  Fig.1(e) &  Fig.1(f) \\
\hline
 $f_1^{\cal Q}$, SU(2)
                    & ${\cal Q} (1 - \displaystyle\frac{9}{4}   R_\pi)$
                    & $\tau_3 R_\pi^b$
                    & $- \tau_3 \displaystyle\frac{\Delta_\pi}{2 F^2}$
                    & $  \tau_3 \displaystyle\frac{\Delta_\pi}{2 F^2}$
                    & $ \displaystyle\frac{1}{2} ({\cal I} - \tau_3) R_\pi^e$
                    & $\tau_3 R_\pi^f$ \\[2mm]
\hline
 $f_1^Q$, SU(3)     & $Q - \sum\limits_P \beta_P^Q R_P$
                    & $\sum\limits_P \lambda_P^b R_P^b$
                    & $- \sum\limits_P \lambda_P^c
  \displaystyle\frac{\Delta_P}{2 F^2}$
                    & $\sum\limits_P \lambda_P^d
  \displaystyle\frac{\Delta_P}{2 F^2}$
                    & $\sum\limits_P \lambda_P^e R_P^e$
                    & $\sum\limits_P \lambda_P^f R_P^f$ \\[2mm]
\hline
 $f_1 = f_1^{du}$, SU(2)
                    & $1 -  \displaystyle\frac{9}{4} R_\pi$
                    & $2 R_\pi^b$
                    & $- \displaystyle\frac{\Delta_\pi}{F^2}$
                    & $  \displaystyle\frac{\Delta_\pi}{F^2}$
                    & $- R_\pi^e$
                    & $2 R_\pi^f$ \\[2mm]
\hline
 $f_1 = f_1^{du}$, SU(3)  & $1 - \sum\limits_P \beta_P R_P$
                    & $2R_\pi^b + R_K^b$
                    & $- \displaystyle\frac{1}{2 F^2}
( 2 \Delta_\pi + \Delta_K)$
                    & $  \displaystyle\frac{1}{2 F^2}
( 2 \Delta_\pi + \Delta_K)$
                    & $- R_\pi^e + \frac{1}{3} R_\eta^e$
                    & $2 R_\pi^f + R_K^f$ \\[2mm]
\hline
 $f_1^{su}$, SU(3)  & $1 - \sum\limits_P \gamma_P R_P$
                    & $I_{\pi K} + I_{\eta K}$
                    & $-  \displaystyle\frac{3}{8 F^2}
(\Delta_\pi + 2 \Delta_K  + \Delta_\eta)$
                    & $J_{\pi K} + J_{\eta K}$
                    & $-  \displaystyle\frac{2}{3} R_\eta^e$
                    & $ \displaystyle\frac{3}{4}
                      (R_\pi^f + 2 R_K^f + R_\eta^f)$ \\
\hline
\end{tabular}
\end{center}

\vspace*{2cm}

\noindent
{\bf Table 2.} Contribution of different diagrams in Fig.2  
to the isotopic (axial) charge $g_1 \tau_3/2$ [in SU(2)]   
and $g_1 \lambda_3/2$ [in SU(3)], axial coupling ($d \to u$  
flavor transition) $g_1^{du} \tau_{ud}$ [in SU(2)] and
$g_1^{du} \lambda_{ud}$ [in SU(3)] and vector coupling 
($s \to u$ flavor transition) $g_1^{su} \lambda_{us}$.
The contribution of diagram in Fig.2(a) is  multiplied 
by the $Z$--factor.

\begin{center}
\def\arraystretch{2}
\begin{tabular}{|c|c|c|c|c|c|c|}
\hline
 Coupling           &  Fig.2(a) &  Fig.2(b) &  Fig.2(c)
                    &  Fig.2(d) &  Fig.2(e) &  Fig.2(f) \\
\hline
 $g_1 = g_1^{du}$, SU(2)
                    & $g ( 1 -  \displaystyle\frac{9}{4} R_\pi ) $
                    & $4 M_\pi^2 d_{16}^q$
                    & $ -  \displaystyle\frac{g}{F^2} \Delta_\pi$
                    & $T_\pi^d$
                    & $T_\pi^e$
                    & $(c_3^q - 2 c_4^q) T_\pi^f$   \\[2mm]
\hline
 $g_1 = g_1^{du}$, SU(3)
                    & $g ( 1 - \sum\limits_P \beta_P R_P )$
                    & $(2 M_\pi^2 + 4 M_K^2) D_{16}^q$
                    & $- \displaystyle\frac{g}{2 F^2}(2\Delta_\pi + \Delta_K)$

                    & $T_\pi^d - \displaystyle\frac{1}{3} T_\eta^d$
                    & $T_\pi^e + \displaystyle\frac{1}{2} T_K^e$
                    & $(C_3^q - 2 C_4^q) T_\pi^f$ \\
                    &
                    & $+ \displaystyle\frac{2}{3} (M_\pi^2 - M_K^2) D_{17}^q$
                    &
                    &
                    &
                    & $-  C_4^q T_K^f$ \\[2mm]
\hline
 $g_1^{su}$, SU(3)  & $g ( 1 - \sum\limits_P \gamma_P R_P )$
                    & $(2 M_\pi^2 + 4 M_K^2) D_{16}^q$
                    & $ - \displaystyle\frac{3 g}{8 F^2}
                      (\Delta_\pi +  2\Delta_K $
                    & $\displaystyle\frac{2}{3} T_\eta^d$
                    & $\displaystyle\frac{3}{8} (T_\pi^e + 2 T_K^e$
                    & $(C_3^q - C_4^q)T_K^f$ \\
                    &
                    & $+ \displaystyle\frac{1}{3} (M_K^2 - M_\pi^2) D_{17}^q$
                    & $+ \Delta_\eta)$
                    &
                    & $+ T_\eta^e)$
                    & $- \displaystyle\frac{1}{2}
                      C_4^q ( 3 T_\pi^f + T_\eta^f )$ \\[2mm]
\hline
\end{tabular}
\end{center}

\newpage

\vspace*{1cm}

\noindent
\begin{center}
{\bf Table 3.} Semileptonic decay constants of baryons
$g_V^{B_iB_j}$ and $g_A^{B_iB_j}$
\end{center}

\begin{center}
\def\arraystretch{2.5}
\begin{tabular}{|c|c|c|}
\hline
Decay mode & $g_V^{B_iB_j}$ & $g_A^{B_iB_j}$ \\
\hline
$n \to p$          & 1 & $\displaystyle{\frac{5}{3}} \,
g_1 \, I_A = g_A = g_A^{\rm SU_3} \, (1 + \delta_{A_1})$
\\[2mm]
\hline
$\Lambda \to p$
& $- \displaystyle{\sqrt{\frac{3}{2}}} \, f_1^{su} \, I_V^s
=  - \displaystyle\sqrt{\frac{3}{2}} \, (1 + \delta_V)$
& $- \displaystyle\sqrt{\frac{3}{2}} \, g_1^{su} \, I_A^s
=  - \displaystyle\frac{3}{5}\sqrt{\frac{3}{2}}
\, g_A^{\rm SU_3} \, (1 + \delta_{A_2})$
\\[2mm]
\hline
$\Sigma^- \to n$   & $- f_1^{su} \, I_V^s = - (1 + \delta_V) $
                   &
$ \displaystyle\frac{1}{3} \, g_1^{su} \, I_A^s
= \displaystyle\frac{1}{5} \,
g_A^{\rm SU_3} \, (1 + \delta_{A_2})$\\[2mm]
\hline
$\Sigma^- \to \Lambda$  & 0
                        & $ \displaystyle\sqrt{\frac{2}{3}} \, g_1 \, I_A
                         =
\displaystyle\frac{\sqrt{6}}{5} \, g_A =
\displaystyle\frac{\sqrt{6}}{5} \,
g_A^{\rm SU_3} \, (1 + \delta_{A_1})$\\[2mm]
\hline
$\Xi^- \to \Lambda$
& $ \displaystyle\sqrt{\frac{3}{2}} \, f_1^{su} \, I_V^s
=   \displaystyle\sqrt{\frac{3}{2}} \, (1 + \delta_V)$
& $ \displaystyle\sqrt{\frac{1}{6}} \, g_1^{su} \, I_A^s
=   \displaystyle\frac{1}{5} \,\sqrt{\frac{3}{2}} \,
g_A^{\rm SU_3} \, (1 + \delta_{A_2})$\\[2mm]
\hline
$\Xi^- \to \Sigma^0$ & $ \displaystyle\sqrt{\frac{1}{2}} \, f_1^{su} \, I_V^s
= \displaystyle\sqrt{\frac{1}{2}} \, (1 + \delta_V)$
& $ \displaystyle\frac{5}{3\sqrt{2}} \, g_1^{su} \, I_A^s
=   \displaystyle\frac{1}{\sqrt{2}} \,
g_A^{\rm SU_3} \, (1 + \delta_{A_2})$\\[2mm]
\hline
$\Xi^0 \to \Sigma^+$   & $f_1^{su} \, I_V^s =  1 + \delta_V$
& $ \displaystyle\frac{5}{3} \, g_1^{su} \, I_A^s
=   g_A^{\rm SU_3} \, (1 + \delta_{A_2})$\\[2mm]
\hline
\end{tabular}
\end{center}

\vspace*{1cm}

\noindent
\begin{center}
{\bf Table 4.} Numerical results for $g_A^{B_iB_j}$
\end{center}

\begin{center}
\def\arraystretch{2}
\begin{tabular}{|c|c|c|}
\hline
Decay mode & Ref.~\cite{FloresMendieta:2006ei}  & Our results \\
\hline
$n \to p$          & 1.272  & 1.2695 \\[2mm]
\hline
$\Lambda \to p$    & -0.904 & -0.944 \\[2mm]
\hline
$\Sigma^- \to n$   & 0.375  & 0.257  \\[2mm]
\hline
$\Sigma^+ \to \Lambda$  & 0.653 & 0.622 \\[2mm]
\hline
$\Sigma^- \to \Lambda$  & 0.624 & 0.622 \\[2mm]
\hline 
$\Xi^- \to \Lambda$     & 0.139 & 0.315 \\[2mm]
\hline
$\Xi^- \to \Sigma^0$    & 0.869 & 0.908 \\[2mm] 
\hline
$\Xi^0 \to \Sigma^+$    & 1.312 & 1.284 \\[2mm]
\hline
\end{tabular}
\end{center}

\newpage

\noindent
\begin{center}
{\bf Table 5.} Numerical results for the semileptonic decay \\
               widths of hyperons (in units of 10$^6$ s$^{-1}$) 
\end{center}

\begin{center}
\def\arraystretch{2}
\begin{tabular}{|c|c|c|}
\hline
Decay mode & Our results & Data~\cite{Yao:2006px}\\
\hline
$\Lambda \to p e^- \bar\nu_e$      & 3.21 & 3.16$\pm$0.06 \\
\hline
$\Lambda \to p \mu^- \bar\nu_\mu$  & 0.52 & 0.60$\pm$0.13 \\
\hline
$\Sigma^- \to n e^- \bar\nu_e$     & 5.50 & 6.88$\pm$0.24 \\
\hline
$\Sigma^- \to n \mu^- \bar\nu_\mu$ & 2.45 & 3.0$\pm$0.2   \\
\hline
$\Sigma^+ \to \Lambda e^+ \nu_e$  & 0.24 & 0.25$\pm$0.06 \\
\hline
$\Sigma^- \to \Lambda e^- \bar\nu_e$  & 0.40 & 0.39$\pm$0.02 \\
\hline 
$\Xi^- \to \Lambda e^- \bar\nu_e$     & 3.11 & 3.35$\pm$0.37 \\
\hline
$\Xi^- \to \Lambda \mu^- \bar\nu_\mu$ & 0.84 & 2.1$^{+2.1}_{-1.3}$ \\
\hline
$\Xi^- \to \Sigma^0 e^- \bar\nu_e$    & 0.51 & 0.53$\pm$0.10 \\
\hline
$\Xi^- \to \Sigma^0 \mu^- \bar\nu_\mu$ & 0.01 & $<$ 0.05  \\[2mm]
\hline
$\Xi^0 \to \Sigma^+ e^- \bar\nu_e$    & 0.90 & 0.88$\pm$0.04\\
\hline
$\Xi^0 \to \Sigma^+ \mu^- \bar\nu_\mu$ & 0.01 & 0.02$\pm$0.01\\
\hline
\end{tabular}
\end{center}

\end{document}